\documentclass[a4paper,8pt]{article}
\pdfoutput=1 

\usepackage{jheppub} 

\usepackage{graphicx,color}
  \usepackage{bm}
   \usepackage{amsmath}
    \usepackage{amssymb}
     \usepackage{pifont}
      \usepackage{simplewick}

\newcommand{\nn}{\nonumber}

\newcommand{\Tr}{\mathrm{Tr}}

\renewcommand{\(}{\left(}
\renewcommand{\)}{\right)}
\renewcommand{\[}{\left[}
\renewcommand{\]}{\right]}

\newcommand{\fnot}[1]{\not{\! #1}}

\title{Calculation of transverse momentum dependent distributions beyond the leading power}

\author{Valentin Moos and}
\author{Alexey Vladimirov}
\affiliation{Institut f\"ur Theoretische Physik, \\ Universit\"at Regensburg,\\
D-93040 Regensburg, Germany}
\emailAdd{alexey.vladimirov@physik.uni-regensburg.de}

\abstract{
We compute the contribution of twist-2 and twist-3 parton distribution functions to the small-$b$ expansion for transverse momentum dependent (TMD) distributions at all powers of $b$. The computation is done by the twist-decomposition method based on the spinor formalism for all eight quark TMD distributions. The newly computed terms are accompanied by the prefactor $(M^2b^2)^n$ and represent the target-mass corrections to the resummed cross-section. For the first time, a non-trivial expression for the pretzelosity distribution is derived.
}

\begin{document}
\maketitle
\flushbottom

\section{Introduction}
\label{sec:introduction}

Transverse momentum dependent (TMD) distributions extend the parton model, including the transverse motion of hadron's constituents. Any TMD distribution is a function of two dynamical variables $x$ and $b$. The variable $x$ is the fraction of hadron's momentum carried by the parton. The variable $b$ is the transverse distance that is Fourier conjugated to the transverse momentum of parton $k_T$. In the limit $b\to0$, which corresponds to integrated or unobserved transverse momentum, a TMD distribution turns to the corresponding collinear parton distribution functions (PDFs) or fragmentation function (FFs). Technically, this relation, which is also known as the matching between TMD parton distribution functions (TMDPDFs) and PDFs (or TMD fragmentation functions (TMDFFs) and FFs), is received by the operator product expansion, and its leading power term is very well studied. In the present work, we extend this formalism beyond the leading power term and compute matching of TMDPDFs to PDFs of \textit{twist-2 and twist-3 at all powers of} $b^2$-expansion.

The matching of TMDPDFs to PDFs is an important part of the TMD factorization approach.  The review of various aspects of TMD factorization can be found in \cite{Collins:2011zzd,Angeles-Martinez:2015sea,Scimemi:2019mlf}. On the theory side, the matching establishes the connection with the resummation formalism and allows interpolation between TMD factorized cross-section and fixed order computations, see f.i. \cite{Becher:2010tm,Gehrmann:2014yya,Collins:2016hqq}. On the phenomenological side, the matching essentially reduces the parametric freedom for TMD ansatzes. The modern phenomenology of TMD distributions is grounded on the matching relations and demonstrates perfect agreement with the large amount of experimental data \cite{Scimemi:2019cmh,Bacchetta:2019sam}. 

So far, all studies of matching relations were restricted to the leading power term only. The leading power term is the most simple and numerically dominant contribution. Nonetheless, several aspects make the study of power corrections interesting. \textit{First of all,} such a study carries a significant amount of methodological novelty. Indeed, the power corrections are generally considered as a complicated field, and their computation is an interesting theoretical task. In this work, we have computed the whole series of power correction with PDFs of twist-2 and twist-3, which is almost unprecedented. Methodologically, the closest example of similar computation is the computation of kinematic power corrections to Deeply Virtual Compton Scattering (DVCS) made by Braun and Manashov in ref.\cite{Braun:2011dg}. \textit{The second point} of interest is the derivation of the matching relations for polarized TMD distributions. For many TMD distributions already, the leading power matching involves twist-3 functions and requires a non-trivial computation. The computations for different TMDPDFs have been made by different methods in refs.\cite{Boer:2003cm,Ji:2006ub,Kang:2011mr,Kanazawa:2015ajw}. In ref.\cite{Scimemi:2018mmi}, all polarized TMDPDFs were systematically computed in a single scheme, and the agreement with previous computations had been shown. However, all these computations were unable to find the non-trivial matching of the pretzelosity distribution, which is derived for the first time in this work. \textit{The third point} of interest is the comparison of the derived power corrections to the extracted ones. There are many examples, where the part of the power correction proportional to twist-2 PDFs (Wandzura-Wilczek approximation) is numerically dominant \cite{Accardi:2009nv}. However, there are also known cases of opposite behavior. In this work, we demonstrate that TMD distributions belong to the latter case, and for them, the contribution of higher twist PDFs is essential. \textit{The forth point} of interest is the target-mass dependence of TMD distributions. At higher powers of small-$b$ series, the target mass is the only scale that compensates the dimension of $b^n$ for twist-2 and twist-3 distributions. Thus, the here derived corrections are target-mass corrections $\sim (M^2b^2)$. Their knowledge is essential since much of the experimental data is measured on nuclei.

Formally, the matching is obtained by operator product expansion of the transverse momentum dependent operator (defined explicitly in (\ref{def:TMDop})) at small values of $b$. The later has the schematic form
\begin{eqnarray}\label{intro:OPE}
O^{\text{TMD}}(z,b)&=&\mathbb{O}_2(z)+b_\mu \mathbb{O}_3^{\mu}(z)+\frac{b_{\mu_1}b_{\mu_2}}{2} \mathbb{O}_4^{\mu_1\mu_2}(z)+...=\sum_{n=0}^\infty \frac{b_{\mu_1}...b_{\mu_n}}{n!}\mathbb{O}_{n+2}^{\mu_1...\mu_n}(z),
\end{eqnarray}
where $z$ is the distance between fields of the operator along the light-cone. The operators $\mathbb{O}_n(z)$ are light-cone operators with the \textit{collinear twist} $n$. For example, the leading power operator is $\mathbb{O}_2(z)\sim \bar q(z n)[zn,0]q(0)$, where $n$ is the light-cone vector, $[a,b]$ is the straight gauge link, and $q$ is the quark field. Each operator $\mathbb{O}_n$ is the integral convolution of a coefficient function and an actual quantum-field operator. The coefficient functions for $\mathbb{O}_2$ are all known at next-to-leading order (NLO) in $\alpha_s$-expansion \cite{Gutierrez-Reyes:2017glx,Buffing:2017mqm}, and NNLO \cite{Gehrmann:2014yya,Echevarria:2016scs,Gutierrez-Reyes:2018iod,Gutierrez-Reyes:2019rug}. The leading power coefficient function for unpolarized distribution has been recently computed at N$^3$LO \cite{Luo:2019szz,Ebert:2020yqt}. Beyond the leading power the information is sparse. The tree order matching for $\mathbb{O}_3$ has been derived in \cite{Scimemi:2018mmi}, see also \cite{Boer:2003cm,Ji:2006ub,Kang:2011mr,Kanazawa:2015ajw} for particular cases. The only NLO computation for $\mathbb{O}_3$ is made for the Sivers function \cite{Scimemi:2019gge}. In this work, we derive only the tree order matching, ignoring the $\alpha_s$-suppressed terms in the coefficient functions. For that reason, we do not specify the renormalization scales and omit corresponding arguments.

The operators with the \textit{collinear twist} $n$ can be presented as a sum of operators with different \textit{geometrical twists},
\begin{eqnarray}\label{intro:twist-decomposition}
\mathbb{O}_{n}(z)=\sum_{t=2}^{n}C_{t}(z;\{y\})\otimes O_t(\{y\}),
\end{eqnarray}
where we omit indices $\mu_1...\mu_n$ for brevity. For shortness, we call this procedure as \textit{twist-decomposition}. Generally, operators $O_t$ depend on many spatial points, which are parameterized by a set of variables $\{y\}$. They are mapped to the single variable $z$ by the integral convolution $\otimes$. The geometric twist has a strict definition as the ``dimension minus spin'' of the operator. Operators with different geometrical twists have different transformation properties and thus represent independent physical observables. The matrix elements of operators with a given geometrical twist define a self-contained set of PDFs. Such a set of PDFs does not mix with other sets, and their evolution is autonomous. For the introduction to the twist decomposition see e.g.\cite{Jaffe:1996zw,Braun:2003rp}, the review of modern development can be found in \cite{Braun:2011dg}. Therefore, the central task is to derive the twist-decomposition for each operator on the right-hand-side (RHS) of (\ref{intro:OPE}). In turn, the small-$b$ expansion of TMDPDFs in terms of collinear PDFs is received by evaluating the matrix element over the derived operator relation.

In the case of FF, the twist-decomposition operation is not well defined. The main reason is the absence of local expansion for fragmentation operators. In ref.\cite{Balitsky:1990ck} it has been shown that OPE for FF is defined up to terms that satisfy the Laplace equation (for the twist-2 part). Therefore, alternative methods such as differential equations \cite{Balitsky:1990ck}, Feynman diagram correspondences \cite{Kanazawa:2013uia,Echevarria:2015usa,Echevarria:2016scs,Gamberg:2018fwy} and Lorentz invariant relations \cite{Kanazawa:2015ajw}, should be used. For that reason, we do not consider the matching of TMDFFs to FFs in the present work. For an interested reader, we present some discussion on power corrections for TMDFF in appendix \ref{sec:FF}.

\begin{figure}[t]
\begin{center}
\includegraphics[width=0.36\textwidth]{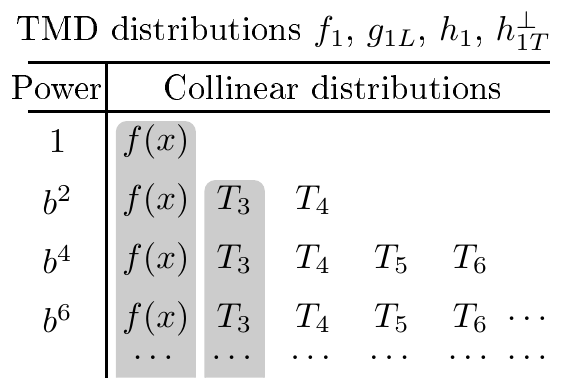}
\hspace{2cm}
\includegraphics[width=0.36\textwidth]{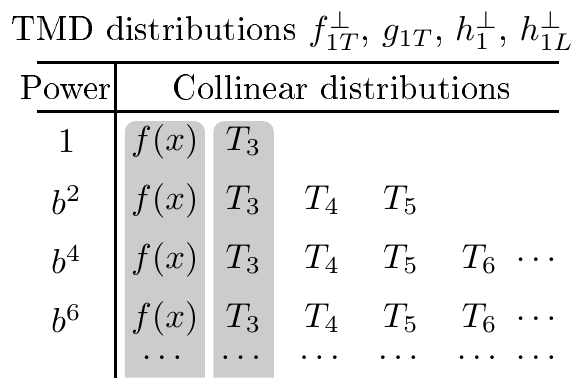}
\caption{\label{fig:powers} Ordering of parton distributions in the small-$b$ series for TMD distributions (\ref{intro:even}) (left) and (\ref{intro:odd}) (right). $f(x)$ denotes the ordinary PDFs. $T_n$ denotes the parton distribution of twist-n, which is generally a function of several variables. The gray boxes designate the terms computed in this work.}

\end{center}
\end{figure}

In this work, we compute only quark TMD distributions, since they are of the prime practical interest. In total, there are eight TMDPDFs at the leading term of the factorization theorem \cite{Mulders:1995dh}. They can be split into two classes with respect to the structure of matching relations. Four distributions, namely $f_1$(unpolarized), $g_{1L}$ (helicity), $h_1$ (transversity) and $h_{1T}^\perp$ (pretzelocity) have contributions of only even collinear twists
\begin{eqnarray}\label{intro:even}
F_{\text{even}}(x,b)&=&f(x)+M^2b^2\sum_{t=2}^4 C_t^{(2)}(x)\otimes T_t+(M^2b^2)^2\sum_{t=2}^6 C_t^{(4)}(x)\otimes T_t+...~,
\end{eqnarray}
where $T_t$ is a collinear distribution of twist-t and $T_2(x)=f(x)$ is the twist-2 PDF. Another four distributions, namely $f_{1T}^\perp$ (Sivers), $g_{1T}$ (worm-gear T), $h_1^\perp$ (Boer-Mulders) and $h_{1L}^\perp$ (worm-gear L), have contributions of only odd \textit{collinear twists}
\begin{eqnarray}\label{intro:odd}
F_{\text{odd}}(x,b)&=&\sum_{t=2}^3 C_t^{(1)}(x)\otimes T_t+M^2b^2\sum_{t=2}^5 C_t^{(3)}(x)\otimes T_t
+...~.
\end{eqnarray}
The graphical representation of these sums is shown in fig.\ref{fig:powers}. The parameter $M$ in (\ref{intro:even},\ref{intro:odd}) is the mass of the hadron, which is inserted such that all coefficient functions $C_t^{(n)}$ are dimensionless. The sums (\ref{intro:even}) and (\ref{intro:odd}) can be reorganized by collecting together distributions of particular twist. For example, equation (\ref{intro:even}) takes the form
\begin{eqnarray}\label{intro:even+}
F_{\text{even}}(x,b)&=&\sum_{n=0}^\infty (M^2b^2)^n C^{(n)}_2(x)\otimes f
\\\nn && +
\sum_{n=1}^\infty (M^2b^2)^n C^{(n)}_3(x)\otimes T_3
+
\sum_{n=1}^\infty (M^2b^2)^n C^{(n)}_4(x)\otimes T_4+...~.
\end{eqnarray}
In the present work, we derive the first and the second terms of this sum for all eight TMD distributions. In fig.\ref{fig:powers}, the shaded areas show the corresponding terms.

To perform this computation, we use the technique inherited from \cite{Braun:2009vc,Braun:2008ia}, where it was used for the analyses of twist-4 operators. The technique is based on the local equivalence of the Lorentz transformation group to SL(2,C) group (so-called spinor formalism) \cite{Dreiner:2008tw}. Within the spinor formalism, the twist-decomposition can be elegantly formulated as the action of a certain spinor-differential operator (see sec.\ref{sec:twist-decomposition}). In ref.\cite{Braun:2011dg} this method has been used to derivate kinematic power corrections $(t/Q^2)$ in deeply-virtual Compton scattering. In contrast to ref.\cite{Braun:2011dg}, TMD operators are essentially non-local. They contain the gauge link along a staple contour. To overcome this difficulty, we introduce a formal local expansion for the TMD operator. To our best knowledge, it is the first time when the series of local operators successfully describes the infinite staple contours. Almost all expressions presented in this article are novel. Only some of them, namely $b^0$- and $b^1$-terms, can be found in literature and agree with it. Additionally, we demonstrate that the series of corrections for the unpolarized TMDPDF $f_1$ can be derived differently using the results of ref.\cite{Balitsky:1987bk} and agrees with it.

The article is organized as follows. In sec.\ref{sec:definitions} we articulate all definitions used in our work. This is an important part because we deal with functions that do not have a common definition such as polarized distributions and PDFs of twist-3. In sec.\ref{sec:spinor} we specify the conventions for the spinor formalism used in our work. Sec.\ref{sec:method} is devoted to the detailed description of the computation method. It is split into three subsections in accordance with three principal stages of the computation:(\ref{intro:OPE}), (\ref{intro:twist-decomposition}) and (\ref{intro:even},\ref{intro:odd}). So, in sec.\ref{sec:TMD-op} the expansion (\ref{intro:OPE}) of the TMD operator in the series of (local) collinear operators is described. In sec.\ref{sec:twist-decomposition} we explain the method of the twist-decomposition and derive it for TMD operators (\ref{intro:twist-decomposition}), with results for twist-3 part presented in the appendix \ref{app:tw3}. The particularities of the computation of matrix elements for twist-decomposed operators are given in sec.\ref{ref:assebling}. The result of the computation and the discussion are given in sec.\ref{sec:results}.

\section{Definitions and conventions}
\label{sec:definitions}

In this work, we operate with eight TMD distributions, three collinear twist-2 distributions and four collinear twist-3 distributions. They essentially depend on the conventions related to P-odd structures, such as Levi-Civita tensor, $\gamma^5$, etc. An additional set of conventions is brought by the spinor formalism which is used for the twist-decomposition.  There is no commonly accepted convention for all these subjects in the literature, e.g. compare conventions in refs.\cite{Bacchetta:2006tn,Braun:2009mi,Jaffe:1996zw,Kang:2011mr,Kanazawa:2015ajw,Sohnius:1985qm,Dreiner:2008tw,Scimemi:2018mmi}. Nicely, the final result is mostly independent on the conventions, because it is the relation between physical distributions. Nonetheless, these conventions play an important role in the intermediate steps. In order to structure the presentation we collect all used definitions and conventions in this section.  

\subsection{Definition of TMD distributions}
\label{sec:TMDdefs}

The light-cone decomposition plays the central role. It is defined by two light-like vectors $n^\mu$ and $\bar n^\mu$ ($n^2=\bar n^2=0$, $(n\bar n)=1$). We use the ordinary notation of vector decomposition
\begin{eqnarray}\label{def:LC-decomposition}
v^\mu=v^+ \bar n^\mu+v^- n^\mu+v_T^\mu,
\end{eqnarray}
where $v^+=(nv)$, $v^-=(\bar n v)$, and $v_T$ is the transverse component $(v_Tn)=(v_T\bar n)=0$. In what follows, the direction $\bar n^\mu$ is associated with the large-component of the hadron momentum $p^\mu$,
\begin{eqnarray}\label{def:pmu}
p^\mu=p^+\bar n^\mu+\frac{n^\mu}{2}\frac{M^2}{p^+},
\end{eqnarray}
where $M$ is the mass of the hadron ($p^2=M^2$). It is important, that the hadron's momentum does not have a transverse component, which gives the physical definition of the transverse plane. The spin of the hadron is parameterized by the spin-vector $S^\mu$, ($S^2=-1$, $(pS)=0$). Its light-cone decomposition is
\begin{eqnarray}\label{def:Smu}
S^\mu=\lambda\frac{p^+}{M}\bar n^\mu-\lambda\frac{M}{2p^+}n^\mu+s_T^\mu,
\end{eqnarray}
where the $\lambda$ is the helicity of the hadron, and $s_T^\mu$ is the transverse component of the spin, $s_T^2=\lambda^2-1$.

The generic quark TMDPDF is defined as
\begin{eqnarray}\label{def:TMDPDF_Qop}
\Phi_{ij}(x,b)&=&\int\frac{dz}{2\pi}e^{-ixz p^+}
\\\nn &&\langle p,S|\bar q_j\(zn+ b\)[z n+ b,\mp\infty n+b][\mp\infty n+b,\mp \infty n]
[\mp\infty n,0] q_i(0)\}|p,S\rangle,
\end{eqnarray}
where the vector $b^\mu$ is a transverse vector, $(bp)=0$. The Wilson lines in the definition (\ref{def:TMDPDF_Qop}) are straight Wilson lines. Rigorously, one should add the T- and anti-T-ordering within the TMD operator. However, for the parton distributions (in contrast to fragmentation functions) it can be safely omitted (see e.g. discussion in \cite{Scimemi:2019gge}). TMDPDFs that appear in different processes have Wilson lines pointing into different direction, which is indicated by $\mp \infty n$ in (\ref{def:TMDPDF_Qop}). So, the TMD distributions which appear in semi-inclusive deep-inelastic scattering (SIDIS) have Wilson lines pointing to $+\infty n$, while in Drell-Yan they point to $-\infty n$. In the following, we distinguish these cases, and the upper sign refers to the Drell-Yan case, whereas the lower sign refers the SIDIS case.

The open indices $(ij)$ of the TMD operator in eq.~(\ref{def:TMDPDF_Qop}) are to be contracted with different gamma-matrices, which we denote generically as $\Gamma$,
\begin{eqnarray}
\Phi^{[\Gamma]}&=& \frac{1}{2}\Tr\(\Phi\Gamma\).
\end{eqnarray}
There are only three Dirac structures that appear in the leading term of the TMD factorization theorem, these are $\Gamma=\{\gamma^+,\gamma^+\gamma_5,i\sigma^{\alpha+}\gamma_5\}$. Here, the index $\alpha$ is transverse and
\begin{eqnarray}
\sigma^{\mu\nu}=\frac{i}{2}(\gamma^\mu\gamma^\nu-\gamma^\nu\gamma^\mu),\qquad \gamma_5=i\gamma^0\gamma^1\gamma^2\gamma^3=\frac{-i}{4!}\epsilon_{\mu\nu\alpha\beta}\gamma^\mu\gamma^\nu\gamma^\alpha\gamma^\beta,
\end{eqnarray}
with $\epsilon^{0123}=-\epsilon_{0123}=1$. In the naive parton model interpretation, these gamma-structures are related to the observation of unpolarized ($\gamma^+$), longitudinally polarized ($\gamma^+\gamma^5$) and transversely polarized ($i\sigma^{\alpha+}_T\gamma^5$) quarks inside the hadron. The standard parameterization of leading TMDPDFs in the position space reads
\begin{eqnarray}\label{param:TMDv}
\Phi^{[\gamma^+]}(x,b)&=&f_1(x,b)+i\epsilon_T^{\mu\nu} b_\mu s_{T\nu} M f_{1T}^\perp(x,b),
\\\label{param:TMDa}
\Phi^{[\gamma^+\gamma_5]}(x, b)&=&\lambda g_{1L}(x,b)+i b_{\mu}s^{\mu}_T M g_{1T}(x,b),
\\\label{param:TMDt}\nn
\Phi^{[i\sigma^{\alpha+}\gamma_5]}(x,b)&=&s_{T}^\alpha h_1(x,b)
 -i\lambda b^{\alpha}M h_{1L}^\perp(x, b)
 \\ &&+i\epsilon_{T}^{\alpha\mu}b_\mu M h_1^\perp(x,b)-\frac{M^2 b^2}{2}\(\frac{ g_{T}^{\alpha\mu}}{2}-\frac{b^\alpha b^\mu}{b^2}\)s_{T\mu} h_{1T}^\perp(x,b).
\end{eqnarray}
 The tensors $g_T^{\mu\nu}$ and $\epsilon_T^{\mu\nu}$ are defined as
\begin{eqnarray}\label{def:epsilon}
g_T^{\mu\nu}=g^{\mu\nu}-n^\mu \bar n^\nu-\bar n^\mu n^\nu,\qquad \epsilon_T^{\mu\nu}=\epsilon^{-+\mu\nu},
\end{eqnarray}
such that $g_T^{11}=g_T^{22}=-1$, $\epsilon^{12}_T=-\epsilon^{21}_T=1$, and the rest components are zero. The definition of the TMDPDFs coincides with the conventional one in \cite{Scimemi:2018mmi,Boer:2011xd,Bacchetta:2006tn}. In the following, we also compare to refs.\cite{Braun:2009mi,Braun:2008ia,Braun:2011dg}, where the definitions of $\epsilon$ and $s_T$ have opposite sign, and ref.\cite{Scimemi:2018mmi}, where the definition of $\epsilon$ and $\epsilon_T$ has opposite sign (so, component-wise the tensor $\epsilon^{\mu\nu}_T$ is the same). TMDPDFs defined in (\ref{param:TMDv}-\ref{param:TMDt}) are dimensionless functions which depend only on the modulus of $b$, but not on the direction. The conventional names for them are (see e.g.\cite{Boer:2011xd,Bacchetta:2006tn}): unpolarized ($f_1$), Sivers ($f_{1T}^\perp$), helicity ($g_{1L}$), worm-gear T ($g_{1T}$), transversity $(h_1)$, worm-gear L ($h_{1L}^\perp$), Boer-Mulders ($h_1^\perp$) and pretzelosity ($h_{1T}^\perp$) distributions. 

The position space representation of TMD distribution is advantageous, because the TMD evolution is multiplicative in the position space. For that reason, the phenomenological studies that incorporate the TMD evolution are made in the position space, for the most recent examples see \cite{Scimemi:2019cmh,Bacchetta:2019sam}. TMD distributions in the momentum space are obtained by the Fourier transformation
\begin{eqnarray}\label{def:p<->b}
\Phi^{[\Gamma]}(x,p_T)=\int \frac{d^2 b}{(2\pi)^2}e^{-i(b p_T)}\Phi^{[\Gamma]}(x, b).
\end{eqnarray}
The transformation rules for particular TMD distributions can be found in refs.\cite{Boer:2011xd,Scimemi:2018mmi}.

\subsection{Definition of collinear distributions}

The collinear distributions of twist-2 are defined as \cite{Jaffe:1996zw}
\begin{eqnarray}\label{def:f1}
\langle p,S|\bar q(zn)[zn,0]\gamma^+ q(0)|p,S\rangle&=& 2p^+\int dx e^{ixzp^+}f_1(x),
\\
\langle p,S|\bar q(zn)[zn,0]\gamma^+\gamma^5 q(0)|p,S\rangle&=& 2\lambda p^+\int dx e^{ixzp^+}g_1(x),
\\
\langle p,S|\bar q(zn)[zn,0]\gamma^+ i\sigma^{\alpha+}\gamma^5 q(0)|p,S\rangle&=& 2s_T^\alpha p^+\int dx e^{ixzp^+}h_1(x),
\end{eqnarray}
where index $\alpha$ is transverse. These distributions are known as unpolarized $(f_1)$, helicity $(g_1)$ and transversity $(h_1)$ PDFs. The variable $x$ belongs to the range $[-1,1]$ and for $x>0$ ($x<0$) PDFs are interpreted as probability densities for (anti-)quarks. 

There is no conventional definition of the collinear distributions of twist-3. Here, we use the definition used in \cite{Scimemi:2018mmi}. We define
\begin{eqnarray}\label{def:T}
&&\langle p,S|g \bar q(z_1n)F^{\mu+}(z_2n)\gamma^+q(z_3n)|p,S\rangle
\\\nn&&\qquad\qquad\qquad
=2 \epsilon_T^{\mu\nu}s_{\nu T} (p^+)^2M \int [dx]e^{-ip^+(z_1x_1+x_2z_2+x_3z_3)}T(x_1,x_2,x_3),
\\\label{def:DeltaT}
&&\langle p,S|g \bar q(z_1n)F^{\mu+}(z_2n)\gamma^+\gamma^5q(z_3n)|p,S\rangle
\\\nn&&\qquad\qquad\qquad
=2i s^\mu_{T} (p^+)^2M \int [dx]e^{-ip^+(z_1x_1+x_2z_2+x_3z_3)}\Delta T(x_1,x_2,x_3),
\\\label{def:deltaTg}
&& \langle p,S|g \bar q(z_1n)F^{\mu+}(z_2n)i\sigma^{\alpha+}\gamma^5q(z_3n)|p,S\rangle
\\\nn&&\qquad\qquad\qquad=
2 \epsilon_T^{\mu\alpha} (p^+)^2M \int [dx]e^{-ip^+(z_1x_1+x_2z_2+x_3z_3)}\delta T_\epsilon(x_1,x_2,x_3)
\\\nn && \qquad\qquad\qquad\qquad+2 i\lambda g_T^{\mu\alpha} (p^+)^2M \int [dx]e^{-ip^+(z_1x_1+x_2z_2+x_3z_3)}\delta T_g(x_1,x_2,x_3),
\end{eqnarray}
where we omit Wilson lines that connect the fields in operators for brevity. The definition of the distributions $T$ and $\Delta T$ coincides with \cite{Scimemi:2018xaf} and \cite{Braun:2009mi}, taking into account the difference in conventions for the $\epsilon$-tensor (explained after (\ref{def:epsilon})). In ref.\cite{Scimemi:2018xaf} one can also find comparison with other definitions. The integration measure $[dx]$ is defined as
\begin{eqnarray}
\int [dx]=\int_{-1}^1 dx_1dx_2dx_3 \delta(x_1+x_2+x_3).
\end{eqnarray}
The delta-function in this measure reflects the independence of the matrix element on the global position $(z_1+z_2+z_3)$ of the field operator. Due to the delta-function in the measure, the distributions of twist-3 effectively depends on only two variables. Nonetheless, it is convenient to keep all three variables $x_{1,2,3}$ as independent. It reveals the symmetric properties \cite{Scimemi:2018mmi}
\begin{eqnarray}\label{coll:sym1}
T(x_1,x_2,x_3)&=&T(-x_3,-x_2,-x_1),
\qquad
\Delta T(x_1,x_2,x_3)=-\Delta T(-x_3,-x_2,-x_1),
\\\label{coll:sym2}
\delta T_\epsilon(x_1,x_2,x_3)&=&\delta T_\epsilon(-x_3,-x_2,-x_1),
\qquad
\delta T_g(x_1,x_2,x_3)=-\delta T_g(-x_3,-x_2,-x_1).
\end{eqnarray}
Each range of $x_{1,2,3}\lessgtr 0$ has a specific partonic interpretation \cite{Jaffe:1996zw}.

\subsection{Spinor formalism}
\label{sec:spinor}

The twist-decomposition for local operators consists in the decomposition of tensors with many indices into irreducible representations of the Lorentz group SO(3,1).  This procedure is greatly simplified in the spinor formalism. The spinor formalism is used in many parts of quantum field theory, for a review see \cite{Sohnius:1985qm,Dreiner:2008tw}. Here we remind only of the essential properties and introduce conventions that are necessary for the current work. 

The spinor formalism is grounded on the local isomorphism of the Lorentz group to the group of complex unimodular matrices SL(2,C). The isomorphism is realized by the map of a four-vector to a hermitian matrix by the rule 
\begin{eqnarray}
x_{\alpha\dot \alpha}=x_\mu\sigma_{\alpha\dot \alpha}^\mu,
\end{eqnarray}
where $\sigma^\mu=\{1,\sigma^1,\sigma^2,\sigma^3\}$ with $\sigma^i$ being the Pauli matrix. The scalar product of any two vectors is $x_\mu y^\mu= x_{\alpha \dot\alpha}y^{\dot\alpha \alpha}/2$. In the spinor formulation one must distinguish dotted and undotted indices because they are related to conjugated representations, $(u_\alpha)^*=\bar u_{\dot \alpha}$. The scalar product of two spinors is defined as
\begin{eqnarray}\label{spinor:SP}
(uv)=-u^\alpha v^\beta \epsilon_{\alpha\beta}=-(vu),\qquad (\bar u\bar v)=-\bar u_{\dot \alpha}\bar v_{\dot \beta}\epsilon^{\dot \alpha\dot \beta}=-(\bar v\bar u),
\end{eqnarray}
where $\epsilon_{12}=-\epsilon_{\dot 1 \dot 2}=1$ as in ref.\cite{Braun:2011dg,Braun:2009vc,Braun:2008ia}.

The light-like vectors $n$ and $\bar n$ in the spinor formalism can be written as
\begin{eqnarray}\label{spinor:nn}
n_{\alpha\dot\alpha}=\lambda_\alpha\bar \lambda_{\dot \alpha},\qquad
\bar n_{\alpha\dot\alpha}=\mu_\alpha\bar \mu_{\dot \alpha},
\end{eqnarray}
where $\lambda$ and $\mu$ are independent spinors normalized as $(\bar \lambda\bar \mu)(\mu\lambda)=2$. The spinors $\lambda$ and $\mu$ form the basis, which can be used to decompose any tensor. In particular, the decomposition (\ref{def:LC-decomposition}) for an arbitrary four-vector is
\begin{eqnarray}\label{spinor:x-decompose}
x_{\alpha\dot \alpha}=\lambda_{\alpha}\bar \lambda_{\dot \alpha} x^-
+\mu_{\alpha}\bar \mu_{\dot \alpha} x^+- \lambda_{\alpha}\bar \mu_{\dot \alpha}x_T - \mu_{\alpha}\bar \lambda_{\dot \alpha}\bar x_T,
\end{eqnarray}
where $x_T$ and $\bar x_T$ are transverse components, $-2x_T\bar x_T=g_T^{\mu\nu}x_\mu x_\nu<0$. 

The Dirac bi-spinors are written as composition of two-component spinors
\begin{eqnarray}\label{spinor:quark}
\bar q=\(\chi^\beta,\bar \psi_{\dot \alpha}\),\qquad q=\(\begin{array}{c}\psi_\alpha \\ \bar \chi^{\dot \beta}
\end{array}\).
\end{eqnarray}
The decomposition of these spinors in the basis (\ref{spinor:nn}) is
\begin{eqnarray}\label{spinor:quark2}
\psi_{\alpha}=\frac{\lambda_\alpha\psi_--\mu_\alpha \psi_+}{(\mu\lambda)},
\qquad
\bar\psi_{\dot\alpha}=\frac{\bar \lambda_{\dot \alpha}\bar \psi_--\bar \mu_{\dot\alpha} \bar \psi_+}{(\bar \lambda\bar\mu)},
\end{eqnarray}
where $\psi_+=(\lambda\psi)$, $\psi_-=(\mu \psi)$, etc.  In the similar way we write down the decomposition of the gluon-strength tensor,
\begin{eqnarray}
F_{\alpha\dot \alpha,\beta\dot\beta}=2(f_{\alpha\beta}\epsilon_{\dot\alpha\dot\beta}-\epsilon_{\alpha\beta}\bar f_{\dot\alpha\dot\beta}),
\end{eqnarray}
where $f_{\alpha\beta}$ and $\bar f_{\dot\alpha\dot \beta}$ are symmetric tensors, $\bar f=f^\dagger$. In our computation we face only the gluon strength-tensor with one index transverse, and another contracted with the vector $n$. The decomposition of such tensor is
\begin{eqnarray}\label{spinor:f++}
F_{\alpha\dot \alpha,+}=\lambda_{\alpha}\bar \lambda_{\dot\alpha}\(\frac{f_{+-}}{(\mu\lambda)}+\frac{\bar f_{+-}}{(\bar \lambda\bar \mu)}\)
-\mu_{\alpha}\bar \lambda_{\dot\alpha}\frac{f_{++}}{(\mu\lambda)}
-\lambda_{\alpha}\bar \mu_{\dot\alpha}\frac{\bar f_{++}}{(\bar \lambda\bar \mu)},
\end{eqnarray}
where $f_{++}=f_{\alpha\beta}\lambda^\alpha\lambda^{\beta}$, $f_{+-}=f_{\alpha\beta}\lambda^\alpha\mu^{\beta}$, etc. The first term in (\ref{spinor:f++}) corresponds to $F_{-+}$ components, whereas the last two terms describe $F_{\mu+}$ with $\mu$ being transverse index.

Using the definitions (\ref{spinor:quark}) and (\ref{spinor:quark2}) we write down the decompositions for the bi-spinor combinations. They are
\begin{eqnarray}\label{chiral-even}
&&\bar q \gamma^+q=\bar \psi_+\psi_++\chi_+\bar \chi_+,\qquad
\bar q \gamma^+\gamma^5q=-\bar \psi_+\psi_++\chi_+\bar \chi_+,
\\\label{chiral-odd}
&&\bar q i\sigma^{(\alpha\dot \alpha)+}\gamma^5 q=-2\(\frac{\mu^\alpha\bar \lambda^{\dot\alpha}}{(\mu\lambda)}\chi_+\psi_++\frac{\lambda^\alpha\bar \mu^{\dot\alpha}}{(\bar\lambda\bar\mu)}\bar\psi_+\bar \chi_+\),
\end{eqnarray}
where the order of the fields on LHS and RHS is preserved. Let us note that only ``plus'' components of quark fields appear in these decompositions. It is not accidental, but is the part of definition for the leading power TMD distributions. The components $\psi_+$, $\chi_+$, etc, are known as ``good'' components of quark field in contrast to ``bad'' components ($\psi_-$, $\chi_-$, etc) \cite{Jaffe:1996zw}. The operators build only from good components (including also good components of the gluon field $f_{++}$ and $\bar f_{++}$) are called \textit{quasi-partonic} operators. Their geometrical twist coincides with the collinear twist. All operators of twist-2 and twist-3 can be expressed as quasi-partonic operators with the help of EOMs \cite{Balitsky:1987bk}.

The last ingredient needed for our computation are the equations of motion (EOMs) for the quark field \,$\fnot \!\!Dq=0$ (for massless quarks). In the spinor notation EOMs are $D^{\dot\alpha\alpha}\psi_\alpha=0$, $\chi^\alpha\overleftarrow{D}_{\alpha\dot \alpha}=0$, similar for other spinors. Contracting these equations with basis spinors one receives EOMs for particular components. For our purposes we need the following EOMs
\begin{eqnarray}\label{EOMs}
D_{\lambda\bar \lambda}\psi_-=D_{\mu\bar \lambda}\psi_+,\qquad
D_{\lambda\bar \lambda}\bar\chi_-=D_{\lambda \bar\mu}\bar\chi_+,
\\
\bar \psi_-\overleftarrow{D_{\lambda\bar \lambda}}=\bar \psi_+\overleftarrow{D_{\lambda\bar\mu}},
\qquad
\chi_-\overleftarrow{D_{\lambda\bar \lambda}}=\chi_+\overleftarrow{D_{\mu\bar\lambda}},
\end{eqnarray}
where $D_{ab}=D_{\alpha\dot\alpha}a^\alpha b^{\dot\alpha}$.

\section{Twist-decomposition for TMD operators}
\label{sec:method}

In this section, we present details of the twist-decomposition procedure. The base of the method is taken from refs.\cite{Braun:2009vc,Braun:2008ia}, to which we refer for extended details and the theory foundation. There are three principal steps of the computation:
\begin{itemize}
\item[1.] The operator is presented as a series of local operators.
\item[2.] Local operators are sorted by irreducible representation of Lorentz group (twists), and simplified using EOMs.
\item[3.] The series of operators with the same twists are summed back into the non-local form.
\end{itemize}
This is a rather traditional approach for twist-decomposition. Examples of such consideration for collinear operators can be found in refs.\cite{Jaffe:1996zw,Braun:2009vc,Braun:2008ia,Balitsky:1987bk,Geyer:1999uq,Belitsky:2000vx}. For TMD operators each of these steps has a certain particularity. Let us list these particularities, and explain the methods that were used to resolve them:
\begin{itemize}
\item[1.] The TMD operator has infinite Wilson lines. Therefore, it is not possible to present it as the series of local operators directly. We regularize the TMD operator by truncation of the length of Wilson lines by the parameter $L$. Then TMD operators can be presented as a limit of triple series of local operators (\ref{TMD=triple-sum}).
\item[2.] The local operators for TMDPDFs have three sets of indices $(s,n,t)$. They correspond to the number of light-cone derivatives acting on $\bar q(s)$, the number of transverse derivatives ($n$) and the number of light-cone derivatives acting on $q(t$). Such structure somewhat complicates the twist-decomposition algebra, in comparison to the case of local operators that describe (Mellin moments of) collinear distributions, where all indices are alike.   To simplify the twist-decomposition procedure we use the method introduced in refs.\cite{Braun:2008ia,Braun:2009vc}, which is based on the spinor formalism. This method allows to perform the twist-decomposition for the most general operators using only differential operations (compare to refs.\cite{Scimemi:2018mmi,Ball:1998sk} where off-light-cone generalizations of operators and integral equations are used, ref.\cite{Balitsky:1987bk} where the differential equation are used, refs.\cite{Geyer:1999uq,Belitsky:2000vx} where an explicit procedure of index symmetrization is performed).
\item[3.] The summation of the series is made for the matrix-elements, i.e. for distributions. It simplifies certain steps of computation, and helps to resolve potential ambiguities in the limit $L\to\infty$.
\end{itemize}
The following sections give details for each step in the order. The results of the computation are collected in the sec.\ref{sec:results}. We stress that such an approach is suitable only for TMDPDFs, but does not apply for TMDFF. The discussion for TMDFFs case is given in sec.\ref{sec:FF}.

\subsection{TMD operator as a series of local operators}
\label{sec:TMD-op}

Let us introduce the TMD operator in the form
\begin{eqnarray}\label{def:TMDop}
O_{\text{TMD}}^{\Gamma}(z,b)=\bar q(zn+b)[zn+b,\mp\infty n+b][\mp\infty b+b,\mp\infty n]\Gamma [\mp\infty n,0]q(0).
\end{eqnarray}
The upper(lower) sign corresponds to Drell-Yan(SIDIS) induced TMD. The transverse gauge link $[\mp\infty n+b,\mp\infty n]$ ensures the explicit gauge invariance of the operator.

At the tree-order quantum fields can be treated as classical fields, and thus the small-$b$ expansion is an ordinary Taylor expansion. It is convenient to write it in the form (\ref{intro:OPE})
\begin{eqnarray}\label{taylor-series}
O_{\text{TMD}}^{\Gamma}(z,b)&=&\sum_{n=0}^\infty \frac{b^{\mu_1}...b^{\mu_n}}{n!}\mathbb{O}^\Gamma_{\mu_1...\mu_n}(z),
\end{eqnarray}
with
\begin{eqnarray}\label{def:OO_n}
\mathbb{O}^\Gamma_{\mu_1...\mu_n}(z)=\bar q(zn)[zn,-\infty n] \overleftarrow{D_{\mu_1}}...\overleftarrow{D_{\mu_n}}\Gamma [-\infty n,0]q(0),
\end{eqnarray}
where $D_\mu$ is the covariant derivative
\begin{eqnarray}
\overleftarrow{D_{\mu}}=\partial_\mu+ig A_\mu,\qquad \overrightarrow{D_{\mu}}=\partial_\mu-ig A_\mu.
\end{eqnarray}
The operators on RHS of (\ref{taylor-series}) have collinear twist $n+2$, which follows from their dimension. At the same time, these operators do not have a definite geometrical twist, and therefore, their matrix element is a complicated composition of collinear distributions with different properties. Our goal is to perform the twist-decomposition and express these operators in terms of operators with definite geometrical twist. 

In contrast to collinear operators, the TMD operator (\ref{def:TMDop}) spans the infinite range along the light cone. It is the most famous feature of TMD operators, and it leads to many physical effects, such as rapidity divergences \cite{Vladimirov:2017ksc}, double-scale nature of TMD evolution \cite{Scimemi:2018xaf} and the sign-change of P-odd distributions \cite{Collins:2002kn}. In the $b\to0$ limit, the infinite Wilson lines are partially compensated, due to the transitivity property of Wilson lines. For example, at $n=0$ the operator (\ref{def:OO_n}) simplifies to
\begin{eqnarray}\label{tw-2-expample1}
\mathbb{O}^\Gamma(z)=\bar q(zn)[zn,-\infty n]\Gamma [-\infty n,0]q(0)=\bar q(zn)[zn,0]\Gamma q(0).
\end{eqnarray}
However, already at $n=1$ the infinities enter the expressions,
\begin{eqnarray}\label{tw-3-example1}
\mathbb{O}^\Gamma_\mu(z)
&=&
\bar q(z)\overleftarrow{D_{\mu}}[zn,0]q(0)
+ig \int_{\mp\infty}^{z} d\sigma \bar q(zn)[zn,\tau n]F_{\mu+}(\tau n)[\tau n,0]q(0),
\end{eqnarray}
where $F_{\mu\nu}=ig^{-1}[D_\mu,D_\nu]$ is the gluon-strength tensor. The operators on the RHS of these formulas are ordinary collinear operators. The infinite size of the TMD operator is reflected in the limit of integration in the last term of (\ref{tw-3-example1}). In the form like (\ref{tw-3-example1}), the twist-decomposition of operators is straightforward, although algebraically complicated \cite{Scimemi:2018mmi}. The main reason of the complication is that each next order of expansion introduces new classes of operators, for instance at $n=2$ the operators like $\bar q(zn)F_{+\mu}(\tau_1n)F_{+\nu}(\tau_2n)q(0)$ and $\bar q(zn)[D_\nu F_{\mu+}(\tau  n)]q(0)$ appear. Each of these new classes requires individual investigation, and therefore, this approach is ineffective. 

To avoid these complications we operate directly with the operators (\ref{taylor-series}) with the help of the following formal procedure. We introduce the regularized operator,
\begin{eqnarray}\label{def:TMDop-L}
O_{\text{TMD}}^{\Gamma}(z,b;L)=\bar q(zn+b)[zn+b, L n+b][ L n+b, L n]\Gamma [L n,0]q(0).
\end{eqnarray}
This operator turns to the TMD operator in the limit $L\to \mp\infty$. Note, that the same regularization also regularizes rapidity divergences and can be used used to derive the non-perturbative definition of the Collins-Soper kernel \cite{Vladimirov:2020umg}. In the regularized form the operator (\ref{def:OO_n}) is
\begin{eqnarray}
\mathbb{O}^\Gamma_{\mu_1...\mu_n}(z;L)=\bar q(zn)[zn,L n] \overleftarrow{D_{\mu_1}}...\overleftarrow{D_{\mu_n}}\Gamma [L n,0]q(0).
\end{eqnarray}
This operator can be written as the formal expansion,
\begin{eqnarray}
\mathbb{O}^\Gamma_{\mu_1...\mu_n}(z;L)=\sum_{s,t=0}^\infty \frac{z_1^sz_2^t}{s!t!} \bar q \overleftarrow{D_+}^s\overleftarrow{D_{\mu_1}}...\overleftarrow{D_{\mu_n}}
\Gamma \overrightarrow{D_+}^tq(L n),
\end{eqnarray}
where 
\begin{eqnarray}\label{def:z1z2}
z_1=z-L,\qquad z_2=-L.
\end{eqnarray}
In this way the TMD operator (\ref{def:TMDop}) is presented as a triple sum 
\begin{eqnarray}\label{TMD=triple-sum}
O_{\text{TMD}}^{\Gamma}(z;b)=\lim_{L\to\mp\infty}\sum_{s,t,n=0}^\infty \frac{z_1^s}{s!}\frac{z_2^t}{t!}\frac{1}{n!}O^\Gamma_{s,n,t}(Ln),
\end{eqnarray}
with
\begin{eqnarray}\label{def:Osnt}
O^\Gamma_{s,n,t}&=& \bar q \overleftarrow{D_+}^s (b\cdot \overleftarrow{D})^n \Gamma \overrightarrow{D_+}^t q.
\end{eqnarray}
The RHS of this expression is suitable for the twist-decomposition procedure.

The limit $L\to\mp\infty$ must be taken with caution. In particular, the summation over $s$ and $t$ must be done before the limiting operation, and the result of the summation should be presented in the form that does not allow any ambiguity. The significant simplification of the summation procedure comes from the possibility to neglect the total derivative terms. The matrix element of a total-derivative operator is proportional to the momentum transfer,
\begin{eqnarray}
\langle p_1|\partial_\mu O|p_2\rangle=i(p_1-p_2)_\mu \langle p_1|O|p_2\rangle.
\end{eqnarray}
Consequently, the total derivative operators do not contribute to TMDPDFs, and we neglect such terms in the following. After elimination of the total derivative contribution the result of summation is independent on $L$ in many cases. The simplest example is $n=0$ case, where
\begin{eqnarray}\label{example-sum1}
\sum_{s,t=0}^\infty \frac{z_1^s}{s!}\frac{z_2^t}{t!}O^\Gamma_{s,0,t}
&=&\sum_{s,t=0}^\infty \frac{z_1^s}{s!}\frac{(-z_2)^t}{t!}O^\Gamma_{s+t,0,0}+\text{total der.}
\\\nn &=& \sum_{s=0}^\infty \frac{(z_1-z_2)^s}{s!}O^\Gamma_{s,0,0}+\text{total der.}~,
\end{eqnarray}
with $z_1-z_2=z$ being independent on $L$. Only in the cases of Sivers and Boer-Mulders functions the limit $L\to\mp\infty$ is not so straightforward. Let us also note that the direction of the limit is the only difference between Drell-Yan $(L\to-\infty)$ and SIDIS $(L\to+\infty)$ cases in the representation (\ref{TMD=triple-sum}).

\subsection{Twist decomposition in the spinor formalism}
\label{sec:twist-decomposition}

The twist-decomposition for the local operators (\ref{def:Osnt}) is equivalent to the decomposition into irreducible representation of Lorentz group. The highest spin representation (completely symmetric and traceless in all vector indices) corresponds to the twist-2 term. The next representation (anti-symmetric for one pair of indices, and symmetric and traceless for all other vector indices) corresponds to twist-3 term. Despite the twist-decomposition is a straightforward operation, it is algebraically complicated, especially for the operators with many transverse indices (for which one has to subtract traces). Additional complication is caused by EOMs, which could relate operators. In refs.\cite{Braun:2009vc,Braun:2008ia} it was observed that the decomposition of higher-indices tensors over irreducible representations is simpler in the spinor representation. The main simplification comes from the anti-symmetry of the scalar product (\ref{spinor:SP}). Due to it, any symmetric tensor (in the spinor space) is automatically traceless. The irreducible representations in SL(2,C) group are obtained by simple symmetrization or anti-symmetrization of spinor indices. This operation can be presented as a differential operator, what essentially simplifies the algebra. Here we present this methods in a slightly modified form, which makes its application more explicit.

Let us consider an operator where all open indices are contracted with basis spinors:
\begin{eqnarray}\label{decomposition:OLambda}
\mathbb{O}_{\Lambda}=\Lambda_1^{\alpha...\beta}(\lambda,\mu)\Lambda_2^{\dot\alpha...\dot\beta}(\bar \lambda,\bar \mu)\mathbb{O}_{\alpha...\beta;\dot\alpha...\dot\beta},
\end{eqnarray}
where $\Lambda_{1,2}$ are monomials of basis spinors, for example (\ref{decomposition:O-snt}). The lowest geometrical twist contribution can be obtained by the symmetrization of dotted and undotted indices (independently). It can be done for the operator, or for the tensor $\Lambda$. Due to the irreducibility of the representation the symmetry properties of one are mapped to another in the convolution. The symmetrization of the tensor $\Lambda$ can be made by the following operation
\begin{eqnarray}
S_n\Lambda^{\alpha...\beta}(\lambda,\mu)=
\(\frac{\mu\partial}{\partial \lambda}\)^{n}\(\frac{\lambda\partial}{\partial \mu}\)^{n}\Lambda^{\alpha...\beta}(\lambda,\mu),
\end{eqnarray}
where 
\begin{eqnarray}\label{def:md/dl}
\frac{\mu\partial}{\partial \lambda}=\mu^\alpha\frac{\partial}{\partial \lambda^{\alpha}},
\qquad
\frac{\lambda\partial}{\partial \mu}=\lambda^\alpha\frac{\partial}{\partial \mu^{\alpha}},
\end{eqnarray}
and $n$ is the number of spinors $\mu$ in the tensor $\Lambda$. The logic behind the construction is the following. First, the action of derivatives $(\lambda\partial)/(\partial\mu)$ replaces all $\mu$'s by $\lambda$'s. The resulting tensor is automatically symmetric in indices. Next, the derivatives $(\mu\partial)/(\partial\lambda)$ replace entries of $\lambda$ by $\mu$ in the fully symmetric fashion. 

The higher twist representations are build by anti-symmetrizing pairs of indices. So, the anti-symmetrization of a single pair can be done by the operator
\begin{eqnarray}
A_1=(\mu^{\alpha}\lambda^\beta-\lambda^\alpha\mu^\beta)\frac{\partial}{\partial \mu^{\alpha}}\frac{\partial}{\partial \lambda^\beta}=\frac{\mu\partial}{\partial\mu}\frac{\lambda\partial}{\partial\lambda}-\frac{\mu\partial}{\partial\lambda}\frac{\lambda\partial}{\partial\mu}+\frac{\mu\partial}{\partial\mu},
\end{eqnarray}
where $(\mu\partial)/(\partial\mu)$ and $(\lambda\partial)/(\partial\lambda)$ defined similarly to (\ref{def:md/dl}). The symmetrization of $n$ pairs is done by
\begin{eqnarray}
A_n=(\mu^{\alpha_1}\lambda^{\beta_1}-\lambda^{\alpha_1}\mu^{\beta_1})...(\mu^{\alpha_n}\lambda^{\beta_n}-\lambda^{\alpha_n}\mu^{\beta_n})
\frac{\partial}{\partial \mu^{\alpha_1}}\frac{\partial}{\partial \lambda^{\beta_1}}...\frac{\partial}{\partial \mu^{\alpha_n}}\frac{\partial}{\partial \lambda^{\beta_n}}.
\end{eqnarray}
The operators $S$ and $A$ commute, $[A_n,S_k]=0$, for any $n$ and $k$. The complete decomposition of a tensor with $n$ entries of spinor $\mu$ then reads
\begin{eqnarray}\label{decomposition:1}
\Lambda^{\alpha...\beta}(\lambda,\mu)=\sum_{k=0}^n c_{n,k}A_{k} S_{n-k}  \Lambda^{\alpha...\beta}(\lambda,\mu),
\end{eqnarray}
where $c_{n,k}$ are numbers depending on the tensor $\Lambda$. Each term in this sum corresponds to an irreducible representation, and the operators $S_{n-k}A_k$ are projectors onto this representation.

To find the coefficients $c_{n,k}$ of the decompositions (\ref{decomposition:1}) we need to normalize the operators $A_kS_n$, such that $(A_kS_n)^2=A_kS_n$. For example, for the symmetrization operator we compute
\begin{eqnarray}
S_nS_n=n! S_n \prod_{m=0}^{n-1}\(\frac{\lambda\partial}{\partial\lambda}+\frac{\mu\partial}{\partial\mu}-m\).
\end{eqnarray}
The operators $(\lambda\partial)/(\partial\lambda)$ and $(\mu\partial)/(\partial\mu)$ count the number of $\lambda$'s and $\mu$'s in the tensor. Therefore, $S_n$ is indeed the projector to the totally symmetric representation, and the normalization factor for $S_n$ is $(N-n)!/(n!N!)$ where $N$ is the total number of indices of the tensor. In the same way, one can check that $A_nS_k$ are projectors, and find the corresponding normalization factors. For our computation we only need the first two terms of the expansion (\ref{decomposition:1}). They are
\begin{eqnarray}
\label{decomposition:2}
\Lambda^{\alpha...\beta}(\lambda,\mu)=\(\frac{(N-n)!}{n!N!}S_n+\frac{(N-n-1)!(N-1)}{(n-1)!N!}A_1S_{n-1}+...\) \Lambda^{\alpha...\beta}(\lambda,\mu),
\end{eqnarray}
where $N$ is the total number of indices, and $n$ is the number of $\mu$'s in the tensor $\Lambda$.

Using this construction we build the operators that extract a part with the certain geometrical twist from the operator (\ref{decomposition:OLambda})
\begin{eqnarray}
O_\Lambda=\sum_{n=2}^\infty \hat T_n O_\Lambda.
\end{eqnarray}
The projectors to the twist-2 is
\begin{eqnarray}\label{decomposition:T2}
\hat T_2=\(\frac{\mu\partial}{\partial \lambda}\)^{n}\(\frac{\bar\mu\partial}{\partial \bar\lambda}\)^{k}\frac{(N-n)!}{n!N!}\frac{(\bar N-k)!}{k!\bar N!}\(\frac{\lambda\partial}{\partial \mu}\)^{n}\(\frac{\bar\lambda\partial}{\partial \bar\mu}\)^{k},
\end{eqnarray}
where $n$, $k$, $N-n$ and $\bar N-k$ are the numbers of $\mu$, $\bar \mu$, $\lambda$ and $\bar \lambda$ in the operator. The operator that projects the twist-3 has two terms
\begin{eqnarray}\label{decomposition:T3+}
\hat T_3=T^{(\mu\lambda)}_3+\hat T^{(\bar \mu\bar\lambda)}_3,
\end{eqnarray}
where $\hat T^{(\mu\lambda)}_3$ anti-symmetrizes a pair $(\mu,\lambda)$, whereas $\hat T^{(\bar \mu\bar \lambda)}_3$ anti-symmetrizes a pair $(\bar\mu,\bar\lambda)$. Explicitly $\hat T^{(\mu\lambda)}_3$ reads
\begin{eqnarray}\label{decomposition:T3}
&&\hat T^{(\mu\lambda)}_3=
\\\nn
&&
\(\frac{\mu\partial}{\partial \lambda}\)^{n-1}\(\frac{\bar\mu\partial}{\partial \bar\lambda}\)^{k}\frac{(N-n-1)!(N-1)}{(n-1)!N!}\frac{(\bar N-k)!}{k!\bar N!}
\(\frac{\mu\partial}{\partial\mu}\frac{\lambda\partial}{\partial\lambda}-\frac{\mu\partial}{\partial\lambda}\frac{\lambda\partial}{\partial\mu}+\frac{\mu\partial}{\partial\mu}\)
\(\frac{\lambda\partial}{\partial \mu}\)^{n-1}\(\frac{\bar\lambda\partial}{\partial \bar\mu}\)^{k},
\end{eqnarray}
where $n$, $k$, $N-n$ and $\bar N-k$ are the numbers of $\mu$, $\bar \mu$, $\lambda$ and $\bar \lambda$ in the operator. The operator $\hat T^{(\bar\mu\bar \lambda)}_3$ is obtained by the interchange of barred and un-barred variables.

In this way, the twist-decomposition is reduced to purely algebraic manipulations with monomials. One should distinguish the chiral-even (given in (\ref{chiral-even})) and chiral-odd (given in (\ref{chiral-odd})) compositions of spinors, because they have different number of barred and un-barred spinors. Apart of this, the evaluation of all cases is alike. In the following we demonstrate the results of action by $T_2$ and $T_3$ on $O^{\Gamma}_{s,n,t}$. For simplicity of presentation, we replace the indication of the Dirac structure in $O^{\Gamma}_{s,n,t}$, by the indication of the corresponding spinor combination. Additionally, we write operators using spinor notation only. For example
\begin{eqnarray}\label{decomposition:O-snt}
O^{\bar \psi_+\psi_+}_{s,n,t}&=& 
\bar \psi_+ \overleftarrow{D_{\lambda\bar \lambda}}^s (b \overleftarrow{D_{\lambda\bar \mu}}+\bar b \overleftarrow{D_{\mu\bar\lambda}})^n \overrightarrow{D_{\lambda\bar \lambda}}^t \psi_+
=(-1)^n2^{s+t+n}\bar \psi_+ \overleftarrow{D_+}^s (b\cdot \overleftarrow{D})^n \Gamma \overrightarrow{D_+}^t \psi_+,
\end{eqnarray}
where the prefactor is originated from the conventions (\ref{spinor:SP}-\ref{spinor:x-decompose}). Also we eliminate all total-derivative terms without indication.

The twist-2 part of the chiral-even operator is
\begin{eqnarray}\label{decomposition:1+}
&&\hat T_2O^{\bar \psi_+\psi_+}_{s,n,t}=
\\\nn &&\qquad\sum_{k=0}^n
\(\frac{\mu\partial}{\partial \lambda}\)^{n-k}\(\frac{\bar\mu\partial}{\partial \bar\lambda}\)^k \frac{(-1)^tn!}{k!(n-k)!}
\frac{(s+t+k+1)!}{(s+t+n+1)!}\frac{(s+t+n-k+1)!}{(s+t+n+1)!}
b^k\bar b^{n-k} \bar \psi_+ \overleftarrow{D}_{\lambda\bar \lambda}^{s+t+n}\psi_+,
\end{eqnarray}
and the same for ${\bar \psi_+\psi_+}\to {\chi_+\bar\chi_+}$. The twist-2 part of the chiral-odd operator is
\begin{eqnarray}\label{decomposition:2+}
&& \hat T_2O^{\chi_+\psi_+}_{s,n,t}=
\\\nn&&
\qquad\sum_{k=0}^n
\(\frac{\mu\partial}{\partial \lambda}\)^{n-k}\(\frac{\bar\mu\partial}{\partial \bar\lambda}\)^k \frac{(-1)^tn!}{k!(n-k)!}
\frac{(s+t+k+2)!}{(s+t+n+2)!}\frac{(s+t+n-k)!}{(s+t+n)!}
b^k\bar b^{n-k} \chi_+ \overleftarrow{D}_{\lambda\bar \lambda}^{s+t+n}\psi_+,
\end{eqnarray}
and the same for ${\chi_+\psi_+}\to {\bar\psi_+\bar\chi_+}$. In equations (\ref{decomposition:1+},\ref{decomposition:2+}) we leave the derivatives with respect $(\mu\partial)/(\partial \lambda)$ and $(\bar\mu\partial)/(\partial \bar\lambda)$ without evaluation for future convenience. Note, that factor $(-1)^t$ is originated from reversing the derivative $\overrightarrow{D}^t$ to $\overleftarrow{D}^t$ and elimination of total derivatives.

The derivation of the twist-3 part is slightly more cumbersome, due to the reduction procedure to the quasi-partonic form. We remind that the quasi-partonic operators consist only of ``good'' components of fields and $D_{\lambda\bar \lambda}$. All twist-3 operators can be reduced to this form. The reduction procedure is done as follows. After the action of $A_1$ we obtain operators of the form $\bar \psi_+ D_{\lambda\bar \lambda}^ND_{\lambda\bar \mu}D_{\lambda\bar \lambda}^M\psi_+$ and $\bar \psi_- D_{\lambda\bar \lambda}^{N}\psi_+$. Next, we commute the derivative with the transverse index to the quark field, such that it can be replaced by appropriate EOM (\ref{EOMs}). After this procedure all ``bad'' components of quark field cancel, and the twist-3 operator has the quasi-partonic form. The expressions for the twist-3 parts of operators are simple but rather lengthy. For example,
\begin{eqnarray}\label{decomposition:3}
\hat T_3^{(\bar \mu\bar\lambda)}O_{s,n,t}^{\bar \psi_+\psi_+}&=&
2ig\sum_{k=1}^n 
\(\frac{\mu\partial}{\partial \lambda}\)^{n-k}
\(\frac{\bar\mu\partial}{\partial \bar\lambda}\)^{k-1}
\\\nn && 
 \frac{(-1)^t(n-1)!}{(k-1)!(n-k)!}
\frac{(s+t+n-k)!(s+t+n)}{(s+t+n+1)!}\frac{(s+t+k+1)!}{(s+t+n+1)!} 
\\\nn && 
b^{k}\bar b^{n-k} (\bar \lambda\bar \mu)\Bigg\{
(s+t+n+1)n\sum_{m=0}^{s-2}+(s+t+n+1)\sum_{m=s-1}^{s+n-2}(s+n-m-1)
\\\nn &&
-n\sum_{m=0}^{s+t+n-2}(s+t+n-m-1)\Bigg\}
\bar \psi_+ \overleftarrow{D}_{\lambda\bar \lambda}^{m}f_{++}\overleftarrow{D}_{\lambda\bar \lambda}^{s+t+n-m-2}\psi_+.
\end{eqnarray}
Here, the summations over $m$ are originated from the commutation procedure. Other spinor combinations and action of $\hat T_3^{(\mu\lambda)}$ differ from this example by $\pm 1$ terms in the factorial factors and summation limits. For completeness these expressions are listed in the appendix \ref{app:tw3}. 

\subsection{Assembling the final result}
\label{ref:assebling}

The last step of the computation is to sum the operators over $s$ and $t$ to the non-local form and perform the limit $L\to \infty$. This procedure can be done directly in terms of operators. However, the resulting expression has an overcomplicated form, because the most part of the expression vanishes on the level of matrix element due to symmetry relations (\ref{coll:sym1},\ref{coll:sym2}). To avoid these complications we first compute the matrix element and then sum the expression. Conveniently, the evaluation of the matrix elements can be done before we apply the derivatives  $(\mu\partial)/(\partial \lambda)$ and $(\bar \mu\partial)/(\partial \bar \lambda)$, which then act on the parameterization of the matrix element. The process is identical for all distributions. Here we demonstrate the computation for the case $\Gamma=\gamma^+$ which contains unpolarized and Sivers TMD distributions. The peculiarities of the computation for other cases are discussed at the end of the section.

The matrix element for the twist-2 operator with $\Gamma=\gamma^+$ is expressed in the terms of unpolarized PDF (\ref{def:f1}) as (\ref{chiral-even})
\begin{eqnarray}\label{assembling:6}
\langle p,S|\bar \psi_+\overleftarrow{D}_{\lambda\bar \lambda}^N\psi_++\chi_+\overleftarrow{D}_{\lambda\bar \lambda}^N\bar\chi_+|p,S\rangle=
i^Np_{\lambda\bar \lambda}^{N+1}\int dx x^N f_1(x).
\end{eqnarray}
Substituting it into (\ref{decomposition:1+}), we observe that the derivatives $(\mu\partial)/(\partial \lambda)$ and $(\bar \mu\partial)/(\partial \bar \lambda)$ now could act only on $p_{\lambda\bar \lambda}$. At the same time, the vector $p^\mu$ does not have a transverse part (by definition), i.e. $p_{\lambda\bar \mu}=p_{\mu\bar \lambda}=0$, and thus only the terms with $n=2k$ are non-zero:
\begin{eqnarray}\label{assembling:3}
\(\frac{\mu\partial}{\partial \lambda}\)^{n-k}\(\frac{\bar\mu\partial}{\partial \bar\lambda}\)^kp_{\lambda\bar \lambda}^{s+t+n+1}
=\delta_{n,2k}\frac{(s+t+n+1)!k!}{(s+t+n-k+1)!}p_{\lambda\bar \lambda}^{s+t+n-k+1}p_{\mu\bar \mu}^{k}.
\end{eqnarray}
Now, we can pass to the vector notation, and the expression for the matrix element takes the form
\begin{eqnarray}\label{assembling:1}
&&\langle p,S|\hat T_2 O^{[\gamma^+]}(z,b)|p,S\rangle=
\\\nn &&
\qquad
p^+ \lim_{L\to \mp\infty} \sum_{n=0}^\infty \frac{1}{n!}\(\frac{b^2M^2}{4}\)^n\sum_{s,t=0}^\infty \frac{z_1^sz_2^t(ip^+)^{s+t}}{s!t!}\frac{(s+t+n+1)!}{(s+t+2n+1)!}\int dx x^{s+t+2n}f_1(x),
\end{eqnarray}
where we used that $b\bar b=-b^2/2$, $p_{\lambda\bar \lambda}=2p^+$ and $p_{\mu\bar \mu}=2p^-=M^2/p^+$. Presenting the factorial factor by the  beta-function (for $n>0$) we evaluate the sum over $s$ and $t$. The summation produces the exponent $\exp(i(z_1-z_2)zp^+)$ that is independent on $L$ due to (\ref{def:z1z2}). Therefore, in this case the limit $L\to \pm \infty$ is trivial. The summed expression reads
\begin{eqnarray}\label{assembling:2}
&&\langle p,S|T_2 O^{[\gamma^+]}(z,b)|p,S\rangle=
\\&&\nn \qquad
p^+ \int dx f_1(x) e^{izp_+x}
+
\int_0^1 du \int dx  \sum_{n=1}^\infty \frac{u^{n+1}\bar u^{n-1}x^{2n}}{n!(n-1)!}\(\frac{b^2M^2}{4}\)^n e^{iuzp_+x}f_1(x),
\end{eqnarray}
where $\bar u=1-u$. Finally, we make the inverse Fourier transformation and compare the result to the parameterization (\ref{param:TMDv}). We find that the twist-2 part of the operator contributes only to the unpolarized TMD distribution. The final expression is convenient to present in the form of Mellin convolution (\ref{def:Mellin})
\begin{eqnarray}
f_1(x,b)=f_1(x)+\sum_{n=1}^\infty \int_0^1 du \int dy \frac{\delta(x-uy)}{n!(n-1)!}\(\frac{\bar u}{u}\)^{n-1}\(\frac{x^2 b^2M^2}{4}\)^n f_1(y)+\mathcal{O}(\alpha_s,\text{tw4}).
\end{eqnarray}
This is the complete part of the small-$b$ expansion for $f_1(x,b)$ that is matches twist-2 PDFs. The corrections to this expression contain PDFs of higher twist, starting from twist-4 (the absence of twist-3 part is demonstrated later). Also each term in this sum receives the perturbative corrections. All these corrections are indicated by $\mathcal{O}(\alpha_s,\text{tw4})$.

The computation of the twist-3 part follows the same pattern, but involves more terms. We use the matrix elements (\ref{def:T},\ref{def:DeltaT}), and present them in the form of double moments
\begin{eqnarray}
&&\langle p,S|g\bar \psi_+\overleftarrow{D}^N_{\lambda\bar \lambda}f_{++}\overleftarrow{D}^M_{\lambda\bar \lambda}\psi_+|p,S\rangle=
\\\nn &&\qquad
(-i)^{N}i^M p_{\lambda\bar \lambda}^{N+M+2}\frac{-iS_{\lambda\bar \mu}}{2(\bar \lambda\bar \mu)}\int [dx] x_1^N x_3^M \(T(x_1,x_2,x_3)+\Delta T(x_1,x_2,x_3)\),
\end{eqnarray}
and similarly for other spinor combinations. Now, the derivatives $(\mu\partial)/(\partial \lambda)$ and $(\bar \mu\partial)/(\partial \bar \lambda)$ also act on the vector $S$, and thus the expression analogous to (\ref{assembling:3}) contains two terms. One with $n=2k-1$ that is proportional to transverse part of $S$, and another one with $n=2k-2$ that is proportional to $S_{\mu\bar \mu}=-\lambda M/p^+$. The parts proportional to $\lambda$ add up to zero for $\Gamma=\gamma^+$, however, in the case of $\Gamma=\gamma^+\gamma^5$ they produce the twist-3 terms in the helicity TMD distribution $g_{1L}$. After these operations the summation over $m$ reduces to geometric progressions. The resulting expression is rather lengthy
\begin{eqnarray}
&&\langle p,S|\hat T_3 O^{[\gamma^+]}(z,b)|p,S\rangle=Mp^+(b_\mu \epsilon_T^{\mu\nu}S_\nu)\lim_{L\to\mp\infty}
\int[dx]\sum_{n=0}^\infty \frac{1}{(2n+1)n!}\(\frac{b^2M^2}{4}\)^n
\\\nn &&\qquad\qquad \sum_{s,t=0}^\infty \frac{z_1^sz_2^t(ip^+)^{s+t}}{s!t!}\frac{(s+t+n)!}{(s+t+2n+2)!}\frac{(s+t+2n+1)(s+t+n+2)}{(s+t+2n+2)}\Bigg\{
\\\nn &&\qquad\qquad (2n+1)\[(-x_1)^sx_1^{t+2n+1}\(2-\frac{x_2}{x_1}\)+(-x_3)^tx_3^{s+2n+1}\(2-\frac{x_2}{x_3}\)\]\frac{T(x_1,x_2,x_3)}{x_2^2}
\\\nn &&\qquad\qquad -(2n+1)\[(-x_1)^sx_1^{t+2n}-(-x_3)^tx_3^{s+2n}\]\frac{\Delta T(x_1,x_2,x_3)}{x_2}
\\\nn &&\qquad\qquad -(s+t+2n+2)(-x_1)^s(-x_3)^t(x_1^{2n+1}+x_3^{2n+1})\frac{T(x_1,x_2,x_3)}{x_2^2}\Bigg\},
\end{eqnarray}
where we have used that $x_1+x_2+x_3=0$ for simplifications. This expression is very representative and demonstrates many features of the computation. Notably, there are two types of contributions relative to the summation over $s$ and $t$. The regular ones that are in the third and forth lines, and the irregular one that is in the last line.

The regular contributions have a general form of (see also (\ref{example-sum1},\ref{assembling:1}))
\begin{eqnarray}
\sum_{s,t=0}^\infty \frac{z_1^s (-z_2)^t}{s!t!}f(s+t)=\sum_{r=0}^\infty \frac{(z_1-z_2)^r}{r!}f(r),
\end{eqnarray}
and are explicitly independent on $L$, since $z_1-z_2=z$ (\ref{def:z1z2}). For regular contributions, the limit $L\to\mp\infty$ is trivial, and the resulting operators are spatially compact. The final expression has ordinary form with twist-3 distributions, see eqns.(\ref{result:g1L},\ref{result:g1T},\ref{result:h1},\ref{result:h1Lperp},\ref{result:h1Tperp}). In the current example with $\Gamma=\gamma^+$, these term do not appear in the final expression, due to symmetry properties (\ref{coll:sym1}). Indeed, the change of variables $\{x_1,x_2,x_3\}\to\{-x_3,-x_2,-x_1\}$ flips the common sign of the prefactors in square brackets, but leaves $T/x_2^2$ and $\Delta T/x_2$ unchanged. Thus, the contribution of the third and the fourth lines vanishes.

The evaluation of irregular terms requires special attention. The sums over $s$ and $t$ lead to the following expression
\begin{eqnarray}\label{assebling:5}
&&\langle p,S|\hat T_3 O^{[\gamma^+]}(z,b)|p,S\rangle=
\\&&\nn\qquad\qquad 
-Mp^+(b_\mu \epsilon_T^{\mu\nu}S_\nu)\lim_{L\to\mp\infty} \int_0^1 du\int[dx]
\sum_{n=0}^\infty\(\frac{b^2M^2}{4}\)^n e^{-iup^+(z_1x_1+z_3x_3)}
\\\nn &&\qquad\qquad \frac{(u\bar u)^n}{(2n+1)n!}\(\delta_{n,0}\delta(\bar u)+ \frac{n}{(n-1)!}\frac{1+(n-1)u+u^2}{\bar u}\)
(x_1^{2n+1}+x_3^{2n+1})\frac{T(x_1,x_2,x_3)}{x_2^2}.
\end{eqnarray}
The limit $L\to\mp\infty$ is ill-defined, because $(z_1x_1+z_3x_3)=(zx_1+Lx_2)$. To resolve this ambiguity, we use the symmetry of the integration measure $[dx]$ and transfer the variable $L$ to the limits of the integration
\begin{eqnarray}\label{assebling:4}
&&\int [dx]e^{-iup^+(z_1x_1+z_3x_3)}
=(-ip^+u)\int[dx]x_2\,e^{-ip_+uzx_1} \int_{-L+z}^{L}d\tau \frac{e^{-ip_+u \tau x_2}}{2}.
\end{eqnarray}
Now, the limit can be taken, and the last integral in (\ref{assebling:4}) is the delta-function
\begin{eqnarray}
\lim_{L\to\mp\infty}(-ip^+u)\int_{-L+z}^{L}d\tau \frac{e^{-ip_+u \tau x_2}}{2}=\mp i\pi \delta(x_2).
\end{eqnarray}
It is important that the sign of this expression depends on the sign of the limiting direction. This is how the famous sign-flip for the Sivers function appears in this computation. To integrate over $x_2$, we extract all entries of $x_2$ in (\ref{assebling:4}) with the help of relation
\begin{eqnarray}
\frac{x_1^{2n+1}+x_3^{2n+1}}{x_2}=-\sum_{m=0}^n (-x_1x_3)^mx_2^{2n-2m}\frac{(2n+1)(2n-m)!}{m!(2n-2m+1)!}.
\end{eqnarray}
Clearly, only the term with $m=0$ contributes. Performing the Fourier transformation and comparing it to (\ref{param:TMDv}) we conclude that it is the contribution to the Sivers function, which reads
\begin{eqnarray}
f_{1T}^\perp(x,b)&=&\pm\pi\Big[T(-x,0,x)
+
\int_0^1 du \int dy \sum_{n=1}^\infty\(\frac{x^2b^2M^2}{4}\)^n
\\\nn && 
 \frac{\delta(x-uy)}{(n-1)!(n+1)!}\(\frac{\bar u}{u}\)^n \frac{1+(n-1)u+u^2}{\bar u}T(-y,0,y)
+\mathcal{O}(\alpha_s,\text{tw4})\Big].
\end{eqnarray}
This is the matching of the Sivers function at all orders of $b^2$-expansion to collinear distributions of twist-2 (which is null) and twist-3. Miraculously, only the Qui-Sterman function $T(-x,0,x)$ \cite{Qiu:1991pp} contributes to this expression. As expected, the sign of the Sivers function depends on the direction of the gauge contour. 

The computation of other Lorentz structures is done in the same way. The differences to the demonstrated computation are immaterial. The resulting expression are not as simple as for $\Gamma=\gamma^+$ case. In particular, generally the distributions have entries of both twist-2 and twist-3 PDFs. It happens due to non-vanishing contributions with derivatives of the vector $S_\mu$. For example, for $\Gamma=\gamma^+\gamma^5$ the analog of (\ref{assembling:6}) has $S_{\lambda\bar \lambda}$ which after the differentiation (\ref{assembling:3}) produces the transverse components of $S$. These terms result into twist-2 PDFs within worm-gear function. In all P-even cases only regular terms contribute, whereas P-odd cases (i.e. Sivers and Boer-Mulders functions) are given by only irregular terms. The results of our computation together with additional comments are presented in the next section.

\section{Results and discussion}
\label{sec:results}

Following the method derived in the previous section we routinely computed all (quark) TMD distributions. Here is the list of expressions. The chiral-even distributions are
\begin{eqnarray}\label{result:f1}
f_1(x,b)&=&f_1(x)+\sum_{n=1}^\infty \int_0^1 du \int dy\frac{\delta(x-uy)}{n!(n-1)!}\(\frac{x^2b^2M^2}{4}\)^n\(\frac{\bar u}{u}\)^{n-1}f_1(y),
\\\label{result:f1Tperp}
f_{1T}^\perp(x,b)&=&\pm\pi\Big[T(-x,0,x)+\sum_{n=1}^\infty  \(\frac{x^2b^2M^2}{4}\)^n \int_0^1 du \int dy\frac{\delta(x-uy)}{(n+1)!(n-1)!}
\\\nn && \qquad \qquad \times \(\frac{\bar u}{u}\)^{n}\frac{1+(n-1)u+u^2}{\bar u}T(-y,0,y)\Big],
\\\label{result:g1L}
g_{1L}(x,b)&=&g_1(x)+\sum_{n=1}^\infty  \(\frac{x^2b^2M^2}{4}\)^{n}\int_0^1 du \int dy\, \frac{\delta(x-uy)}{n!(n-1)!}
\\\nn &&
\qquad \qquad \times 
\(\frac{\bar u}{u}\)^{n-1}\Big[
\(1-2n\frac{\bar u}{u}\Psi_n(u)\)g_1(y)-2\frac{\bar u}{u}\(1-\frac{\bar u}{u}\Psi_{n+1}(u)\)\mathcal{T}_g(y)\Big],
\\\label{result:g1T}
g_{1T}(x,b)&=&x \int_0^1 du \int dy \delta(x-uy)\Big\{
g_1(y)+\mathcal{T}_g(y)+\sum_{n=1}^\infty\(\frac{x^2b^2M^2}{4}\)^n\frac{1}{n!(n-1)!}
\\\nn &&
\qquad \qquad  \times  
\(\frac{\bar u}{u}\)^{n}\Big[\Psi_n(u)g_1(y)+\(\frac{1}{n}+\frac{1}{n+1}\(\frac{\bar u}{u}\)^2\Psi_{n+2}(u)\)\mathcal{T}_g(y)\Big].
\end{eqnarray}
The chiral-odd distributions are
\begin{eqnarray}\label{result:h1}
h_1(x,b)&=& h_1(x)+\sum_{n=1}^\infty \(\frac{x^2b^2 M^2}{4}\)^n\int_0^1du \int dy  \frac{\delta(x-uy)}{n!(n-1)!}
\\\nn &&  \qquad \times \(\frac{\bar u}{u}\)^n\Big[\(\frac{u}{\bar u}+\frac{\bar u}{n+1}\)h_1(y)-
x^2\((n+u)-\frac{(n-1)n}{u}\Psi_n(u)\)
\mathcal{T}_h(y)\Big],
\\\label{result:h1perp}
h_1^\perp(x,b)&=&\mp i\pi \Big[\delta T_\epsilon(-x,0,x)
\\\nn &&\qquad +\sum_{n=1}^\infty\(\frac{x^2 b^2M^2}{4}\)^n\int dy\int_0^1 du
 \frac{\delta(x-uy)}{n!(n-1)!} \(\frac{\bar u}{u}\)^{n-1} \delta T_{\epsilon}(-y,0,y)\Big],
\\\label{result:h1Lperp}
h_{1L}^\perp(x,b)&=& -x\sum_{n=0}^\infty 
\(\frac{x^2 b^2 M^2}{4}\)^n \int_0^1du \int dy \frac{\delta(x-uy)}{n!n!}
\\\nn &&
\qquad\times \(\frac{\bar u}{u}\)^n\Big[\frac{u+n}{n+1}h_1(y)+\(u-n\frac{\bar u}{u}+n^2 \frac{\bar u}{u^2}\Psi_{n+1}(u)\)
\mathcal{T}_h(y)\Big],
\\
\label{result:h1Tperp}
h_{1T}^\perp(x,b)&=&
-x^2
\sum_{n=0}^\infty \(\frac{x^2b^2M^2}{4}\)^n\int_0^1 du\int dy\frac{\delta(x-uy)}{n!(n+1)!}
\\\nn &&\qquad \times \(\frac{\bar u}{u}\)^{n+1}
\((n+u+1)-\frac{n(n+1)}{u}\Psi_{n+1}(u)\)
\mathcal{T}_h(y).
\end{eqnarray}
In these expression $u=1-u$, and
\begin{eqnarray}
\Psi_n(u)=\Phi\(\frac{u-1}{u},1,n\)=(-1)^n\(\frac{u}{\bar u}\)^n\ln u-\sum_{k=1}^{n-1}\frac{(-1)^k}{n-k}\(\frac{u}{\bar u}\)^k,
\end{eqnarray}
where $\Phi$ is the Lerch function. The twist-3 PDFs are conveniently gathered into the following combinations
\begin{eqnarray}
\mathcal{T}_g(x)&=&\int[dx] \Big[\delta(x-x_3)\(\frac{\Delta T(x_{1,2,3})}{x_2^2}+\frac{T(x_{1,2,3})-\Delta T(x_{1,2,3})}{2x_2 x_3}\)
\\\nn &&\qquad\qquad\qquad\qquad-\delta(x+x_1)\(\frac{\Delta T(x_{1,2,3})}{x_2^2}-\frac{T(x_{1,2,3})+\Delta T(x_{1,2,3})}{2x_2 x_3}\)
\Big],
\\
\mathcal{T}_h(x)&=&\int[dx] \Big[\delta(x-x_3)\(\frac{\delta T_g(x_{1,2,3})}{x_2^2}-\frac{\delta T_g(x_{1,2,3})}{x_2 x_3}\)
\\\nn && \qquad\qquad\qquad\qquad\qquad\qquad
-\delta(x+x_1)\(\frac{\delta T_g(x_{1,2,3})}{x_2^2}-\frac{\Delta T_g(x_{1,2,3})}{x_2 x_3}\)
\Big],
\end{eqnarray}
where we use the shorthand notation $(x_{1,2,3})=(x_1,x_2,x_3)$. For Sivers and Boer-Mulders functions the upper(lower) sign corresponds to the Drell-Yan (SIDIS) configuration. The collinear distributions of the twist-2 and twist-3 are defined in (\ref{def:f1} - \ref{def:deltaTg}). All these expressions have the form of the Mellin convolution. For numerical computation it can be rewritten as
\begin{eqnarray}\label{def:Mellin}
[G\otimes F](x)=\int_0^1 du \int dy \delta(x-uy)G(u)F(y)=
\left\{\begin{array}{ll}
\int_{x}^1 \frac{dy}{y}G\(\frac{x}{y}\)F(y),& x>0 ,
\\
\int_{-x}^1 \frac{dy}{y}G\(\frac{-x}{y}\)F(-y),& x<0.
\end{array}\right.
\end{eqnarray}
The point $u=1$ is regular for all integrals (\ref{result:f1}-\ref{result:h1Tperp}). The parameter of the expansion $x^2M^2b^2/4$ is negative, due to $b^2<0$. In the following we discuss particularities of each TMD.

\textit{Unpolarized TMD $f_1$.} The unpolarized TMD $f_1(x,b)$ is given in (\ref{result:f1}). The leading term of this expression is well-known. Moreover, the perturbative corrections for the leading term are know up to $\alpha_s^3$-order \cite{Luo:2019szz,Ebert:2020yqt}. Peculiarly, the contributions proportional to the twist-3 are absent at all powers. So, the next contribution to $f_1$ contains twist-4 PDFs. 

\begin{figure}
\begin{center}
\includegraphics[width=0.45\textwidth]{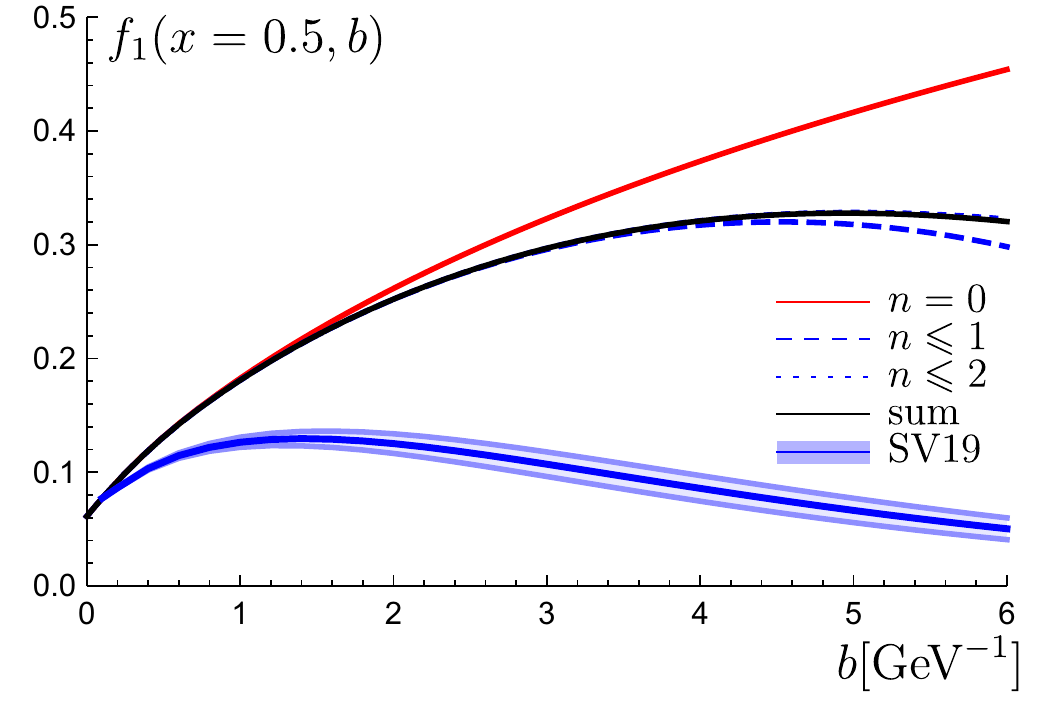}
~~
\includegraphics[width=0.45\textwidth]{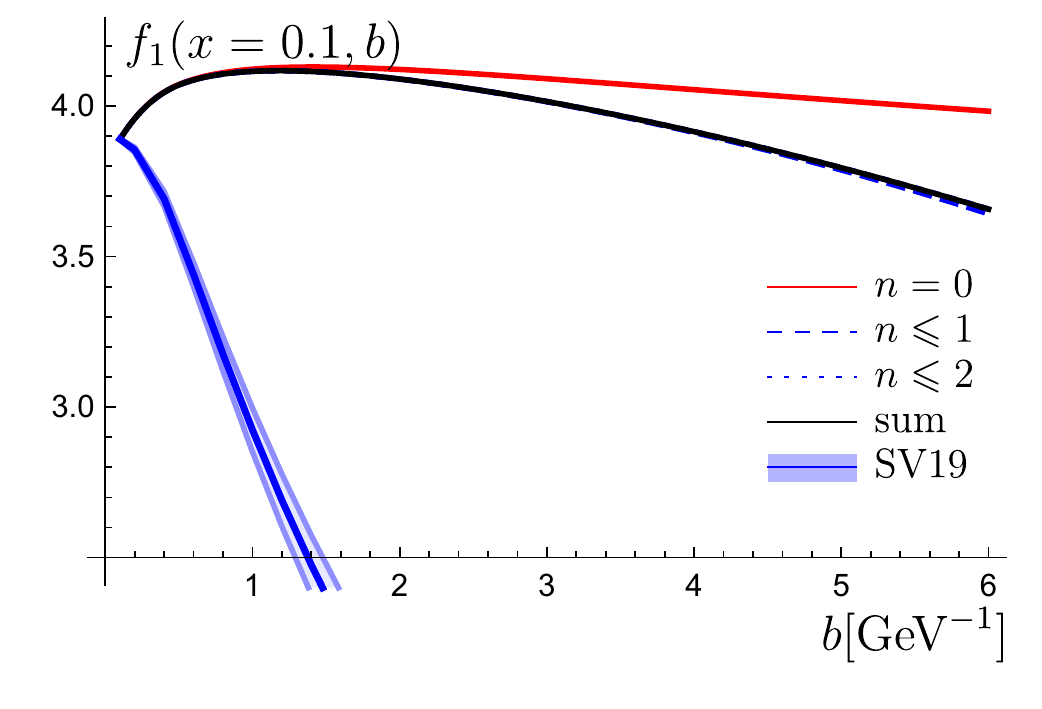}
\caption{\label{fig:f1} Unpolarized TMD (for d-quark) as the function of $b$ at different orders of mass corrections. The curve $n=0$ corresponds to the leading term. The curves $n\leqslant 1,2$ correspond to the partial sums (\ref{result:f1}) up to $n=1,2$. In all cases, the PDF is taken in the convolution with NNLO perturbative coefficient function. The curve labeled as ``sum'' is the summed expression for target-mass corrections (\ref{f1:sum}). The curve with a band is the extraction of non-perturbative  TMD distribution $f_1$ made in ref.\cite{Scimemi:2019cmh}.}

\end{center}
\end{figure}

The series (\ref{result:f1}) can be summed to the Bessel function with the result
\begin{eqnarray}\label{f1:sum}
f_1(x,b)&=&f_1(x)-\frac{xM|b|}{2}\int_0^1 du\int dy \delta(x-uy) \sqrt{\frac{u}{\bar u}} J_1\(xM|b|\sqrt{\frac{\bar u}{u}}\)f_1(y)
\end{eqnarray}
where $|b|=\sqrt{-b^2}$. However, the value of the expansion parameter $(xM|b|/2)^2$ is so small that the numerical difference between the summed series and the first term is vanishing (for non-extreme values of $b$). Moreover, since the cross-section for TMD observables is dominated by small values of $b$ ($b<1$GeV$^{-1}$), the derived corrections are negligible for many experiments (e.g. for the LHC, where typically $x\sim 10^{-2}$). The comparison of (\ref{f1:sum}) truncated series (\ref{result:f1}) is shown in fig.\ref{fig:f1}. Fig.\ref{fig:f1} also shows the unpolarized TMDPDF $f_1(x,b)$ extracted in ref.\cite{Scimemi:2019cmh}. The difference between theoretical and extracted curves demonstrates that the target-mass corrections give negligibly small portion of power corrections. The larger portion of corrections is given by the twist-4 (and higher) distributions. In ref.\cite{Scimemi:2016ffw} it has been demonstrated that infrared renormalon poles for unpolarized TMD $f_1$ are proportional to $(xM^2b^2)^n$, and therefore, higher-twist contributions must be enhanced by at least a power of $x^{-n}$.

The unpolarized distribution is the only case where the expression for target-mass corrections can be compared to the literature. To make the comparison, we mention that the twist-2 operators (at the tree order) are not affected by the direction of the Wilson lines. Therefore, to obtain the matching to the twist-2 operators (at the tree order), one could ignore the staple contour of the TMD operator and connect the quark-fields by the straight gauge link. This operator is the usual QCD string-operator, and the small-$b$ expansion is the ordinary $x^2$-expansion used in DIS. The twist-2 part of such operator at all powers of $x^2$ can be found in eqn.(5.1) of ref.\cite{Balitsky:1987bk}, where it has been derived by means of differential equations. The expression in \cite{Balitsky:1987bk} is given in the coordinate space, and after the Fourier transformation coincides with (\ref{result:f1}). It gives a non-trivial check of the computation method and results.

\textit{Sivers $f_{1T}^\perp$ and Boer-Mulders $h_1^\perp$ functions.} Sivers and Boer-Mulders functions are given in (\ref{result:f1Tperp}) and (\ref{result:h1perp}) correspondingly. Both functions are P-odd, and thus they have a different sign for Drell-Yan and SIDIS configurations \cite{Collins:2002kn}. In our computation, the sign-flip arises due to the order of integration limits in the irregular terms (\ref{assebling:4}). These are the only terms where the infinite size of the Wilson lines (the parameter $L$) plays a role. The leading power matching for the Sivers function is known for a long time \cite{Ji:2006ub,Kang:2011mr}, whereas the leading power matching Boer-Mulders function was derived in \cite{Scimemi:2018mmi}. Our computation agrees with these references.

In the limit $L\to\mp\infty$ the coefficient function reduces to $\delta(x_2)$ and thus Sivers and Boer-Mulders functions are expressed through $T(-x,0,x)$ and $\delta T_\epsilon(-x,0,x)$ at all powers of $b^2$. The function $T(-x,0,x)$ is known as a Qiu-Sterman (QS) function \cite{Qiu:1991pp}, and $\delta T_\epsilon(-x,0,x)$ is the analog of QS function for the chiral-odd operator. Both TMDPDFs are expressed via QS functions at all orders of $b$-expansion. This observation is a non-trivial fact because the QS functions are not autonomous, and mix with the bulk of twist-3 distribution during the evolution \cite{Braun:2009mi}. The NLO expression for the leading power coefficient function \cite{Scimemi:2019gge} contains the non-QS but only in the logarithmic terms (i.e., the terms responsible for the evolution effects), whereas the finite part is proportional to the QS function. Based on this information, we make a conjecture that Sivers and Boer-Mulders functions can be expressed entirely through the QS function, up to logarithmic terms.

\textit{Helicity $g_{1L}$ and transversity $h_1$ TMDs.} The helicity and tranversity TMDPDFs are given in (\ref{result:g1L}) and (\ref{result:h1}) correspondingly. These distributions have leading power matching to the twist-2 distributions. The perturbative coefficients for the leading power matching are known up to NLO \cite{Gutierrez-Reyes:2017glx,Buffing:2017mqm} (for helicity) and NNLO \cite{Gutierrez-Reyes:2018iod} (for transversity). In contrast, to the unpolarized TMD $f_1$ (\ref{result:f1}) distribution $g_{1L}$ and $h_1$ have non-zero contribution of twist-3 PDFs. The expressions given in this work do not account the twist-4 PDFs, and thus already the $n=1$ term is incomplete and contains additional twist-4 contribution (see also fig.\ref{fig:powers}).

\textit{Worm-gear TMDs $g_{1T}$ and $h_{1L}^\perp$.} The worm-gear distributions $g_{1T}$ and $h_{1L}^\perp$ are given in (\ref{result:g1T}) and (\ref{result:h1Lperp}), correspondingly. They have the generic structure of distributions with the leading power matching at collinear twist-3 operator (see fig.\ref{fig:powers}). Both worm-gear TMDPDFs have the factor $x$, which suppresses them at small-$x$. It is interesting to mention that despite both distribution have similar form, TMD $g_{1T}$ is expected to be much larger according to a recent measurement \cite{Airapetian:2020zzo}.  The leading power expressions for both distributions were derived in \cite{Kanazawa:2015ajw} (using Lorentz invariance relations) and \cite{Scimemi:2018mmi} (using the off-light-cone parametrization), and agree with our computation. 

\textit{Pretzelosity TMD $h_{1T}^\perp$.} The expression for the pretzelosity TMD is given in (\ref{result:h1Tperp}). The expression (\ref{result:h1Tperp}) demonstrates a non-trivial value for pretzelosity distribution for the first time. The leading power term for the pretzelosity reads
\begin{eqnarray}\label{pretzelosity:2}
h_{1T}^\perp(x,b)=-x^2\int_0^1 du \int dy \delta(x-uy)\frac{1-u^2}{u}\mathcal{T}_h(y).
\end{eqnarray}
This expression is incomplete, because at the same level of accuracy PDFs of twist-4 could contribute. Thus eqn.(\ref{pretzelosity:2}) is only the Wandzura-Wilchek approximation to the full expression. Note, that unless $\mathcal{T}_h(x)$ has some anomalously strong behavior, (that could happen only at $x\to0$, since $\mathcal{T}_h(1)=0$) the pretzelosity distribution is a very small function. At $x\to1$ the convolution is suppressed by the factor $(1-u^2)$, whereas at $x\to0$ there is a common factor $x^2$. This observation is in a general agreement with the experimental data \cite{Airapetian:2020zzo,Lefky:2014eia}, which is compatible with zero for the pretzelosity-involving observables.

There are no contribution of twist-2 PDFs. The absence of matching to twist-2 PDFs is true for all orders of the perturbation theory that were checked in ref.\cite{Gutierrez-Reyes:2018iod} up to $\alpha_s^2$-order, and discussed in ref.\cite{Chai:2018mwx}. This fact can be proven comparing the following combinations of chiral-odd TMDPDFs
\begin{eqnarray}\label{pretz1}
b_\alpha \Phi^{[i\sigma^{\alpha+}\gamma_5]}(x,b)&=&(bs_T)\(h_1(x,b)+\frac{M^2b^2}{2}h_{1T}^\perp(x,b)\)-i\lambda b^2 Mh_{1L}^\perp(x,b),
\\\label{pretz2}
b_\alpha \epsilon_T^{\alpha\beta} \Phi^{[i\sigma^{\beta+}\gamma_5]}(x,b)&=&(b_\alpha \epsilon^{\alpha\beta} s_\beta)\(h_1(x,b)-\frac{M^2b^2}{2}h_{1T}^\perp(x,b)\)-ib^2 Mh_{1}^\perp(x,b).
\end{eqnarray}
The non-zero pretzelosity appears only if the matrix element $\Phi^{[i\sigma^{\alpha+}\gamma_5]}$ has terms with different parity (so the expressions in (\ref{pretz1},\ref{pretz2}) in brackets are different). At the twist-2 level there is only one PDF, and thus the pretzelosity is null. At the twist-3 level there are pairs of operators with different parity: e.g. $\bar \psi_+f_{++}\bar \chi_+$ and $\bar \psi_+\bar f_{++}\bar \chi_+$ which produce asymmetry in (\ref{pretz1},\ref{pretz2}), because $\langle p,S|\bar \psi_+\bar f_{++}\bar \chi_+|p,S\rangle=0$. Since QCD Lagrangian conserves parity, these statements are generally preserved at any order of perturbative expansion in a properly defined regularization scheme. In the dimension regularization the Levi-Civita tensor is not uniquely defined and thus the symmetry between relations (\ref{pretz1},\ref{pretz2}) could be violated. In this case, one observes the non-zero pretzelosity \cite{Gutierrez-Reyes:2017glx,Gutierrez-Reyes:2018iod} in the $\epsilon$-suppressed terms. However, it is only an artifact of the dimensional regularization, and it must vanish at $\epsilon\to0$. The same statement holds for the quasi-TMDs for which the non-trivial but $\epsilon$-suppressed matching to pretzelocity has been observed in ref.\cite{Ebert:2020gxr} at $\alpha_s$-order. 

\section{Conclusion}
\label{sec:discussion}

\begin{table}
\begin{tabular}{l|c|| c  |c |c || c | c}
	 & 			& Twist of	& Twist-2 		& Twist-3		&	Order of	& \\
Name & Function & leading	& distributions & distributions	&	leading power 		&Ref. \\
	 &			& matching  & in matching	& in matching	&	coef.function  & 
\\\hline\hline
unpolarized & $f_1(x,b)$ & tw-2 & $f_1(x)$ & -- & N$^3$LO ($\alpha_s^3$) & \cite{Luo:2019szz,Ebert:2020yqt}
\\\hline
Sivers & $f_{1T}^\perp(x,b)$ & tw-3 & -- & $T(-x,0,x)$ & NLO ($\alpha_s^1$) & \cite{Scimemi:2019gge}
\\\hline
helicity & $g_{1L}(x,b)$ & tw-2 & $g_1(x)$ & $\mathcal{T}_g(x)$ & NLO ($\alpha_s^1$) & \cite{Gutierrez-Reyes:2017glx,Buffing:2017mqm}
\\\hline
worm-gear T & $g_{1T}(x,b)$ & tw-2/3 & $g_1(x)$ & $\mathcal{T}_g(x)$ & LO ($\alpha_s^0$) & \cite{Kanazawa:2015ajw,Scimemi:2018mmi}
\\\hline\hline
transversity & $h_{1}(x,b)$ & tw-2 & $h_1(x)$ & $\mathcal{T}_h(x)$ & NNLO ($\alpha_s^2$) & \cite{Gutierrez-Reyes:2018iod}
\\\hline
Boer-Mulders & $h_{1}^\perp(x,b)$ & tw-3 & -- & $\delta T_\epsilon(-x,0,x)$ & LO ($\alpha_s^0$) & \cite{Scimemi:2018mmi}
\\\hline
worm-gear L & $h_{1L}^\perp(x,b)$ & tw-2/3 & $h_1(x)$ & $\mathcal{T}_h(x)$ & LO ($\alpha_s^0$) & \cite{Kanazawa:2015ajw,Scimemi:2018mmi}
\\\hline
pretzelosity & $h_{1T}^\perp$ & tw-3/4 & -- & $\mathcal{T}_h(x)$ & LO ($\alpha_s^0$) & eq.(\ref{result:h1Tperp})
\end{tabular}
\caption{\label{tab:summary} Summary of the information on the collinear matching for quark TMDs.}
\end{table}

TMD distributions are related to the collinear distributions in the limit of small-$b$. In the present work, we have studied this relation at all powers of $b$-expansion and derived the contributions with twist-2 and twist-3 PDFs. Our computation includes all eight polarized TMDs. From the perspective of the resummation approach, the computed corrections are target-mass corrections. The summary of the here-derived and known results is presented in table \ref{tab:summary}. It is the first study of the matching between TMDPDF and collinear PDFs beyond the leading power, to our best knowledge. Thus, many results and conclusions made in this paper are novel.

The list of expression for TMDPDFs is presented in sec.\ref{sec:results}. All expressions have the following common structure
\begin{eqnarray}\label{conclusion:1}
F(x,b)=\sum_{n=0}^\infty \(\frac{x^2b^2M^2}{4}\)^n \frac{1}{n!n!}\[\(\frac{\bar u}{u}\)^n G_n(u) \otimes f\](x),
\end{eqnarray}
where $F$ is a TMD distribution, $f$ is a (combination of) collinear distribution, and $\otimes$ is the Mellin convolution. In addition to the power-suppression each term is accompanied by the factor $x^{2n}$. Due to its numerical value of computed correction is almost negligible. The appearance of target-mass corrections in the combination $(x^2M^2)$ justifies the usage of proton TMDPDF for nuclei. For nuclear observables the target-mass is $Z$-times larger, but the measured $x$ is $Z$-times smaller (with $Z$ being the atomic number). Together these factors compensate each other, and thus the nuclear TMDPDF is roughly a nucleon TMDPDF. The knowledge of target-mass corrections is also essential for lattice computations of polarized quasi-TMDPDFs \cite{Vladimirov:2020ofp,Ebert:2020gxr}.

One of the central results of this work is the elaboration of the twist-decomposition-method. The method is taken from \cite{Braun:2008ia,Braun:2009mi} and uses the simplifications of the spinor formalism. In the present study, it is applied at the tree-order, but similarly it can be used together with loop-calculation. Let us note that the computation is made in the position space, which is the only straightforward way to compute such power corrections. It is clear that each term with $n>0$ of the expansion (\ref{conclusion:1}) is power-divergent in the the momentum space. Moreover, the resummed series of power corrections (\ref{f1:sum}) also has a divergent Fourier transform. It demonstrates the well-known fact that at certain $b$, the perturbative series (in any form) fails to describe non-local objects, and truly non-perturbative effects must be accounted.

For the first time, we derive the non-zero matching of pretzelosity distribution to the collinear distributions. Its leading power expression contains twist-3 (given in (\ref{pretzelosity:2})) and twist-4 PDFs (not computed). Previously there were unsuccessful attempts to find a non-zero matching of pretzelosity to twist-2 PDFs \cite{Gutierrez-Reyes:2017glx,Gutierrez-Reyes:2018iod}. In sec.\ref{sec:results} we provide the argumentation of why the pretzelosity does not have a contribution of twist-2 PDFs at all orders of power and $\alpha_s$ corrections.

\section*{Acknowledgements}
We thank Alexander Manashov and Vladimir Braun for numerous discussions and explanations. The research was supported by DFG grant N.430824754 as a part of the Research Unit FOR 2926.

\appendix

\section{Twist-3 part of the operators $O_{s,n,t}$}
\label{app:tw3}

In this appendix, we collect the expressions for twist-3 parts of the operators $O_{s,n,t}$. The definition of operator $\hat T_3$ is given in (\ref{decomposition:T3+},\ref{decomposition:T3}). The definition of the operator $O_{s,n,t}$ is given in (\ref{def:Osnt},\ref{decomposition:O-snt}). The route of derivation is explained in sec.\ref{sec:twist-decomposition}.

The twist-3 part of chiral-even operators are
\begin{eqnarray}
\hat T_3^{(\bar \mu\bar\lambda)}O_{s,n,t}^{\bar \psi_+\psi_+}&=&
2ig\sum_{k=1}^n 
\(\frac{\mu\partial}{\partial \lambda}\)^{n-k}
\(\frac{\bar\mu\partial}{\partial \bar\lambda}\)^{k-1}
\\\nn && \frac{(-1)^t(n-1)!}{(k-1)!(n-k)!}
\frac{(s+t+n-k)!(s+t+n)}{(s+t+n+1)!}\frac{(s+t+k+1)!}{(s+t+n+1)!} 
\\\nn && 
b^{k}\bar b^{n-k} (\bar \lambda\bar \mu)\Bigg\{
(s+t+n+1)n\sum_{m=0}^{s-2}+(s+t+n+1)\sum_{m=s-1}^{s+n-2}(s+n-m-1)
\\\nn &&
-n\sum_{m=0}^{s+t+n-2}(s+t+n-m-1)\Bigg\}
\bar \psi_+ \overleftarrow{D}_{\lambda\bar \lambda}^{m}f_{++}\overleftarrow{D}_{\lambda\bar \lambda}^{s+t+n-m-2}\psi_+
,
\end{eqnarray}
\begin{eqnarray}
\hat T_3^{(\mu\lambda)}O_{s,n,t}^{\bar \psi_+\psi_+}
&=&-2ig
\sum_{k=1}^n
\(\frac{\bar\mu\partial}{\partial \bar\lambda}\)^{n-k}
\(\frac{\mu\partial}{\partial \lambda}\)^{k-1} 
\\\nn &&
\frac{(-1)^{s+n}(n-1)!}{(k-1)!(n-k)!}\frac{(s+t+n-k)!(s+t+n)}{(s+t+n+1)!}\frac{(s+t+k+1)!}{(s+t+n+1)!}
\\\nn && \bar b^{k} b^{n-k}(\mu\lambda)
\Bigg\{
(s+t+n+1)n\sum_{m=0}^{t-2}+(s+t+n+1)\sum_{m=t-1}^{t+n-2}(t+n-m-1)
\\\nn &&
-n\sum_{m=0}^{s+t+n-2}(s+t+n-m-1)\Bigg\}
\bar \psi_+ \overrightarrow{D}_{\lambda\bar \lambda}^{s+t+n-m-2}\bar f_{++}\overrightarrow{D}_{\lambda\bar \lambda}^{m}\psi_+,
\end{eqnarray}
and
\begin{eqnarray}
\hat T_3^{(\bar \mu\bar\lambda)}O_{s,n,t}^{\chi_+\bar \chi_+}&=&\hat T_3^{(\mu\lambda)}O_{s,n,t}^{\bar \psi_+\psi_+}\{\bar \psi_+\psi_+\to\chi_+\bar \chi_+, a\leftrightarrow\bar a\},
\\
\hat T_3^{(\mu\lambda)}O_{s,n,t}^{\chi_+\bar \chi_+}&=&\hat T_3^{(\bar \mu\bar\lambda)}O_{s,n,t}^{\bar \psi_+\psi_+}\{\bar \psi_+\psi_+\to\chi_+\bar \chi_+, a\leftrightarrow\bar a\},
\end{eqnarray}
where $a\leftrightarrow\bar a$ indicates that all barred and unbarred variables should be exchanged, namely $\mu$, $\lambda$, $b$ and $f_{++}$. The twist-3 part of chiral-odd operators have the same general form, but different coefficients in the sum
\begin{eqnarray}
\hat T_3^{(\bar \mu\bar\lambda)}O_{s,n,t}^{\bar \psi_+\bar \chi_+}&=&
2ig\sum_{k=1}^n 
\(\frac{\mu\partial}{\partial \lambda}\)^{n-k}
\(\frac{\bar\mu\partial}{\partial \bar\lambda}\)^{k-1}
\\\nn &&
 \frac{(-1)^t(n-1)!}{(k-1)!(n-k)!}
\frac{(s+t+n-k+1)!(s+t+n+1)}{(s+t+n+2)!}\frac{(s+t+k)!}{(s+t+n)!} 
\\\nn && 
b^{k}\bar b^{n-k} (\bar \lambda\bar \mu)\Bigg\{
(s+t+n+2)n\sum_{m=0}^{s-2}+(s+t+n+2)\sum_{m=s-1}^{s+n-2}(s+n-m-1)
\\\nn &&
-n\sum_{m=0}^{s+t+n-2}(s+t+n-m)\Bigg\}
\bar \psi_+ \overleftarrow{D}_{\lambda\bar \lambda}^{m}f_{++}\overleftarrow{D}_{\lambda\bar \lambda}^{s+t+n-m-2}\bar\chi_+
,
\end{eqnarray}
\begin{eqnarray}
\hat T_3^{(\bar \mu\bar\lambda)}O_{s,n,t}^{\chi_+ \psi_+}&=&
2ig\sum_{k=1}^n 
\(\frac{\mu\partial}{\partial \lambda}\)^{n-k}
\(\frac{\bar\mu\partial}{\partial \bar\lambda}\)^{k-1} 
\\\nn &&
\frac{(-1)^t(n-1)!}{(k-1)!(n-k)!}
\frac{(s+t+n-k+1)!(s+t+n+1)}{(s+t+n+2)!}\frac{(s+t+k)!}{(s+t+n)!} 
\\\nn && 
b^{k}\bar b^{n-k} (\bar \lambda\bar \mu)\Bigg\{
(s+t+n)n\sum_{m=0}^{s-2}+(s+t+n)\sum_{m=s-1}^{s+n-2}(s+n-m-1)
\\\nn &&
-n\sum_{m=0}^{s+t+n-2}(s+t+n-m-1)\Bigg\}
\chi_+ \overleftarrow{D}_{\lambda\bar \lambda}^{m}f_{++}\overleftarrow{D}_{\lambda\bar \lambda}^{s+t+n-m-2}\psi_+
,
\end{eqnarray}
and
\begin{eqnarray}
\hat T_3^{(\mu\lambda)}O_{s,n,t}^{\bar\psi_+\bar \chi_+}&=&-\hat T_3^{(\bar\mu\bar\lambda)}O_{s,n,t}^{\chi_+ \psi_+}\{\chi_+ \psi_+\to\bar\psi_+\bar \chi_+, a\leftrightarrow\bar a\},
\\
\hat T_3^{(\mu\lambda)}O_{s,n,t}^{\chi_+ \psi_+}&=&-\hat T_3^{(\bar\mu\bar\lambda)}O_{s,n,t}^{\bar\psi_+\bar \chi_+}\{\bar\psi_+\bar \chi_+ \to \chi_+ \psi_+, a\leftrightarrow\bar a\}.
\end{eqnarray}

\section{Power corrections for fragmentation functions}
\label{sec:FF}

In the case of TMDFF the matching to the collinear FF is not entirely defined. The reason is the absence of local-operator expansion for FF-type operators. Only indirect methods of twist-decomposition for FF operators are possible, such as differential equations \cite{Balitsky:1990ck}, Feynman diagram correspondences \cite{Kanazawa:2013uia,Echevarria:2015usa,Echevarria:2016scs,Gamberg:2018fwy} and Lorentz invariant relations \cite{Kanazawa:2015ajw}. However, all these methods are ambiguous, and allow an addition boundary contributions. In this appendix we demonstrate the result which appears if we partially ignores these problems. For detailed discussion we refer to \cite{Balitsky:1990ck}.

The quark TMDFF is defined as \cite{Collins:2011zzd}
\begin{eqnarray}\label{def:TMDFF}
&&\Delta_{ij}(x,b)=\frac{\Tr_{\text{color}}}{2xN_c}\int \frac{dz}{2\pi}e^{-izp^+/x}
\\\nn &&\qquad \sum_X \langle 0|[\mp \infty n,0]q_i(0)|h(p,s)+X\rangle\langle h(p,s)+X|\bar q(zn+b)[zn+b,\mp\infty n+b]|0\rangle,
\end{eqnarray}
where we use $x$ for collinear fraction of momentum instead of traditional $z$ to avoid the clash of notation. The Wilson lines are connected to $-\infty$ for SIDIS-like process, and to $+\infty$ for $e^+e^-$-annihilation-like processes. For spinless particles there are only two TMDFFs that contribute to the leading term of the factorization theorem. They read
\begin{eqnarray}
\Delta^{[\gamma^+]}(x,b)=D_1(x,b),\qquad \Delta^{[i\sigma^{\alpha+}\gamma^5]}(x,b)=i\epsilon^{\alpha\mu}_T b_\mu M H_1^\perp(x,b).
\end{eqnarray}
These functions are called unpolarized ($D_1$) and Collins ($H_1^\perp$) functions. In contrast to the TMDPDF operator, the TMDFF operator cannot be presented as a single T-ordered operator. This is the central issue because it prevents application of OPE. The definition of collinear FF of twist-2,
\begin{eqnarray}\label{def:d1}
&& d_1(x)=
\\\nn && x\frac{\Tr_{\text{color}}}{2N_c}\int \frac{dz}{2\pi}e^{-izp^+/x}
\sum_X \langle 0|\frac{\gamma^+}{2}[-\infty n,0]q_i(0)|h(p,s)+X\rangle\langle h(p,s)+X|\bar q(zn)[zn,-\infty n]|0\rangle.
\end{eqnarray}
Note, that collinear FF is defined with relative factor $x^2$ in comparison to TMDFF \cite{Collins:2011zzd}.

In the case of FF-type operators the procedure described in sec.\ref{sec:method} should be modified. The reason is the operator in (\ref{def:TMDFF}) has two parts which could not be joined under a single T-order sign. So, one must distinguish causal (indicated by $(+)$-sign) and anti-causal (indicated by $(-)$-sign) fields which then interact by a cut-propagator. Altogether it is known as Keldysh technique. In Keldysh technique the TMDFF-operator can be written analogously to (\ref{def:TMDop})
\begin{eqnarray}\label{def:TMDFFop}
O_{\text{TMDFF}}^{\Gamma}(z,b)=\bar q^{(-)}(zn+b)[zn+b,\mp\infty n+b]^{(-)}\Gamma [\mp\infty n,0]^{(+)}q^{(+)}(0),
\end{eqnarray}
where $[a,b]^{(\pm)}$ is a Wilson line made with $A^{(\pm)}$. TMDFF operator can be decomposed into the series (\ref{TMD=triple-sum}) with
\begin{eqnarray}\label{def:Osnt-FF}
O^\Gamma_{FF;s,n,t}&=& \bar q^{(-)} \overleftarrow{D^{(-)}_+}^s (b\cdot \overleftarrow{D}^{(-)})^n \Gamma \overrightarrow{D^{(+)}_+}^t q^{(+)}.
\end{eqnarray}
Next, one can perform the twist-decomposition for this operator in complete analogy to the TMDPDF operator. The final expressions for twist-2 part (\ref{decomposition:1+}-\ref{decomposition:2+}) are analogous, with the only replacement $D_{\lambda\bar \lambda}^{s+t+n}\to (D^{(-)}_{\lambda\bar \lambda})^{s+n}(D^{(+)}_{\lambda\bar \lambda})^{t}$. However, the expression for twist-3 part is significantly modified by addition of extra terms with $F_{\mu\nu}^{(+-)}=ig^{-1}[D_\mu^{(+)}D_\nu^{(-)}]$. These terms cannot be written in a ``quasi-partonic'' form. It is an unsolved issue.

The next principal difficulty appears when we combine operators to the non-local form. We have not found a way to perform this procedure on the operator level such that the limit $L\to \pm \infty$ can be taken. Alternatively, we can use the analog of (\ref{assembling:6}) 
\begin{eqnarray}\label{FF:6}
\frac{\Tr_{\text{color}}}{N_c}\sum_X\langle 0|\gamma^+\overrightarrow{D}_+^N q|h(p,s)+X\rangle\langle h(p,s)+X|\bar q \overleftarrow{D}_+^M|0\rangle=
4i^{N+M}p_{\lambda\bar \lambda}^{N+1}\int dx \frac{d_1(x)}{x^{N+M+1}}.
\end{eqnarray}
Let us note that this is not a very well defined expression, because it assumes that LHS is dependent on $N+M$ only, which generally does not hold. Nonetheless, operating in this way we receive the all-order expression for unpolarized TMDFF (compare to (\ref{assembling:6}))
\begin{eqnarray}\label{FF:1}
D_1(x,b)&=&\frac{d_1(x)}{x^2}+\sum_{n=1}^\infty \int_1^\infty du \int dy\frac{\delta(x-uy)}{n!(n-1)!}\(\frac{b^2M^2}{4x^2}\)^n \frac{(u-1)^{n-1}}{u} d_1(y).
\end{eqnarray}
The most notable part here is the ``improper'' range of the integration over $u$. The ambiguity in the definition of OPE for FF, allows us to add any function that satisfies Laplace equation. For the detailed discussion we refer to sec.3 and 4 of ref.\cite{Balitsky:1990ck}, where it is shown that for the unpolarized case a convenient addendum is expression (\ref{FF:1}) with integration range for $u$ extended to $(0,\infty)$. Subtracting this expression we get
\begin{eqnarray}\label{FF:2}
D_1(x,b)&=&\frac{d_1(x)}{x^2}-\sum_{n=1}^\infty \int_0^1 du \int dy\frac{\delta(x-uy)}{n!(n-1)!}\(\frac{b^2M^2}{4x^2}\)^n \frac{(u-1)^{n-1}}{u} d_1(y).
\end{eqnarray}
This expression satisfies the Gribov-Lipatov relation between diagrams of PDF and FF kinematics \cite{Gribov:1972rt}. This expression can be also derived from equations (3.23), (3.24) and (5.17) of ref.\cite{Balitsky:1990ck}, in the same fashion as (\ref{result:f1}) can be derived from \cite{Balitsky:1987bk} (see explanation in sec.\ref{sec:results}). In contrast to TMDPDF case, the target-mass corrections for TMDFF are enhanced by $1/x^2$ factor. It makes them large for baryon FFs. For meson TMDFFs these corrections remains small due to the small mass of mesons.

\bibliographystyle{JHEP}
\bibliography{TMD_bib}

\providecommand{\href}[2]{#2}\begingroup\raggedright\begin{thebibliography}{10}

\bibitem{Collins:2011zzd}
J.~Collins, \emph{{Foundations of perturbative QCD}}, vol.~32.
\newblock Cambridge University Press, 11, 2013.

\bibitem{Angeles-Martinez:2015sea}
R.~Angeles-Martinez et~al., \emph{{Transverse Momentum Dependent (TMD) parton
  distribution functions: status and prospects}},
  \href{http://dx.doi.org/10.5506/APhysPolB.46.2501}{\emph{Acta Phys. Polon. B}
  {\bfseries 46} (2015) 2501--2534},
  [\href{https://arxiv.org/abs/1507.05267}{{\ttfamily 1507.05267}}].

\bibitem{Scimemi:2019mlf}
I.~Scimemi, \emph{{A short review on recent developments in TMD factorization
  and implementation}},
  \href{http://dx.doi.org/10.1155/2019/3142510}{\emph{Adv. High Energy Phys.}
  {\bfseries 2019} (2019) 3142510},
  [\href{https://arxiv.org/abs/1901.08398}{{\ttfamily 1901.08398}}].

\bibitem{Becher:2010tm}
T.~Becher and M.~Neubert, \emph{{{Drell-Yan} Production at Small $q_T$,
  Transverse Parton Distributions and the Collinear Anomaly}},
  \href{http://dx.doi.org/10.1140/epjc/s10052-011-1665-7}{\emph{Eur. Phys. J.
  C} {\bfseries 71} (2011) 1665},
  [\href{https://arxiv.org/abs/1007.4005}{{\ttfamily 1007.4005}}].

\bibitem{Gehrmann:2014yya}
T.~Gehrmann, T.~Luebbert and L.~L. Yang, \emph{{Calculation of the transverse
  parton distribution functions at next-to-next-to-leading order}},
  \href{http://dx.doi.org/10.1007/JHEP06(2014)155}{\emph{JHEP} {\bfseries 06}
  (2014) 155}, [\href{https://arxiv.org/abs/1403.6451}{{\ttfamily 1403.6451}}].

\bibitem{Collins:2016hqq}
J.~Collins, L.~Gamberg, A.~Prokudin, T.~Rogers, N.~Sato and B.~Wang,
  \emph{{Relating Transverse Momentum Dependent and Collinear Factorization
  Theorems in a Generalized Formalism}},
  \href{http://dx.doi.org/10.1103/PhysRevD.94.034014}{\emph{Phys. Rev. D}
  {\bfseries 94} (2016) 034014},
  [\href{https://arxiv.org/abs/1605.00671}{{\ttfamily 1605.00671}}].

\bibitem{Scimemi:2019cmh}
I.~Scimemi and A.~Vladimirov, \emph{{Non-perturbative structure of
  semi-inclusive deep-inelastic and Drell-Yan scattering at small transverse
  momentum}}, \href{http://dx.doi.org/10.1007/JHEP06(2020)137}{\emph{JHEP}
  {\bfseries 06} (2020) 137},
  [\href{https://arxiv.org/abs/1912.06532}{{\ttfamily 1912.06532}}].

\bibitem{Bacchetta:2019sam}
A.~Bacchetta, V.~Bertone, C.~Bissolotti, G.~Bozzi, F.~Delcarro, F.~Piacenza
  et~al., \emph{{Transverse-momentum-dependent parton distributions up to
  N$^3$LL from Drell-Yan data}},
  \href{https://arxiv.org/abs/1912.07550}{{\ttfamily 1912.07550}}.

\bibitem{Braun:2011dg}
V.~Braun and A.~Manashov, \emph{{Operator product expansion in QCD in
  off-forward kinematics: Separation of kinematic and dynamical
  contributions}}, \href{http://dx.doi.org/10.1007/JHEP01(2012)085}{\emph{JHEP}
  {\bfseries 01} (2012) 085},
  [\href{https://arxiv.org/abs/1111.6765}{{\ttfamily 1111.6765}}].

\bibitem{Boer:2003cm}
D.~Boer, P.~Mulders and F.~Pijlman, \emph{{Universality of T odd effects in
  single spin and azimuthal asymmetries}},
  \href{http://dx.doi.org/10.1016/S0550-3213(03)00527-3}{\emph{Nucl. Phys. B}
  {\bfseries 667} (2003) 201--241},
  [\href{https://arxiv.org/abs/hep-ph/0303034}{{\ttfamily hep-ph/0303034}}].

\bibitem{Ji:2006ub}
X.~Ji, J.-W. Qiu, W.~Vogelsang and F.~Yuan, \emph{{A Unified picture for single
  transverse-spin asymmetries in hard processes}},
  \href{http://dx.doi.org/10.1103/PhysRevLett.97.082002}{\emph{Phys. Rev.
  Lett.} {\bfseries 97} (2006) 082002},
  [\href{https://arxiv.org/abs/hep-ph/0602239}{{\ttfamily hep-ph/0602239}}].

\bibitem{Kang:2011mr}
Z.-B. Kang, B.-W. Xiao and F.~Yuan, \emph{{QCD Resummation for Single Spin
  Asymmetries}},
  \href{http://dx.doi.org/10.1103/PhysRevLett.107.152002}{\emph{Phys. Rev.
  Lett.} {\bfseries 107} (2011) 152002},
  [\href{https://arxiv.org/abs/1106.0266}{{\ttfamily 1106.0266}}].

\bibitem{Kanazawa:2015ajw}
K.~Kanazawa, Y.~Koike, A.~Metz, D.~Pitonyak and M.~Schlegel, \emph{{Operator
  Constraints for Twist-3 Functions and Lorentz Invariance Properties of
  Twist-3 Observables}},
  \href{http://dx.doi.org/10.1103/PhysRevD.93.054024}{\emph{Phys. Rev. D}
  {\bfseries 93} (2016) 054024},
  [\href{https://arxiv.org/abs/1512.07233}{{\ttfamily 1512.07233}}].

\bibitem{Scimemi:2018mmi}
I.~Scimemi and A.~Vladimirov, \emph{{Matching of transverse momentum dependent
  distributions at twist-3}},
  \href{http://dx.doi.org/10.1140/epjc/s10052-018-6263-5}{\emph{Eur. Phys. J.
  C} {\bfseries 78} (2018) 802},
  [\href{https://arxiv.org/abs/1804.08148}{{\ttfamily 1804.08148}}].

\bibitem{Accardi:2009nv}
A.~Accardi, A.~Bacchetta and M.~Schlegel, \emph{{What can we learn from the
  breaking of the Wandzura-Wilczek relation?}},
  \href{http://dx.doi.org/10.1063/1.3203299}{\emph{AIP Conf. Proc.} {\bfseries
  1155} (2009) 35--41}, [\href{https://arxiv.org/abs/0905.3118}{{\ttfamily
  0905.3118}}].

\bibitem{Gutierrez-Reyes:2017glx}
D.~Gutiérrez-Reyes, I.~Scimemi and A.~A. Vladimirov, \emph{{Twist-2 matching
  of transverse momentum dependent distributions}},
  \href{http://dx.doi.org/10.1016/j.physletb.2017.03.031}{\emph{Phys. Lett. B}
  {\bfseries 769} (2017) 84--89},
  [\href{https://arxiv.org/abs/1702.06558}{{\ttfamily 1702.06558}}].

\bibitem{Buffing:2017mqm}
M.~G.~A. Buffing, M.~Diehl and T.~Kasemets, \emph{{Transverse momentum in
  double parton scattering: factorisation, evolution and matching}},
  \href{http://dx.doi.org/10.1007/JHEP01(2018)044}{\emph{JHEP} {\bfseries 01}
  (2018) 044}, [\href{https://arxiv.org/abs/1708.03528}{{\ttfamily
  1708.03528}}].

\bibitem{Echevarria:2016scs}
M.~G. Echevarria, I.~Scimemi and A.~Vladimirov, \emph{{Unpolarized Transverse
  Momentum Dependent Parton Distribution and Fragmentation Functions at
  next-to-next-to-leading order}},
  \href{http://dx.doi.org/10.1007/JHEP09(2016)004}{\emph{JHEP} {\bfseries 09}
  (2016) 004}, [\href{https://arxiv.org/abs/1604.07869}{{\ttfamily
  1604.07869}}].

\bibitem{Gutierrez-Reyes:2018iod}
D.~Gutierrez-Reyes, I.~Scimemi and A.~Vladimirov, \emph{{Transverse momentum
  dependent transversely polarized distributions at
  next-to-next-to-leading-order}},
  \href{http://dx.doi.org/10.1007/JHEP07(2018)172}{\emph{JHEP} {\bfseries 07}
  (2018) 172}, [\href{https://arxiv.org/abs/1805.07243}{{\ttfamily
  1805.07243}}].

\bibitem{Gutierrez-Reyes:2019rug}
D.~Gutierrez-Reyes, S.~Leal-Gomez, I.~Scimemi and A.~Vladimirov,
  \emph{{Linearly polarized gluons at next-to-next-to leading order and the
  Higgs transverse momentum distribution}},
  \href{http://dx.doi.org/10.1007/JHEP11(2019)121}{\emph{JHEP} {\bfseries 11}
  (2019) 121}, [\href{https://arxiv.org/abs/1907.03780}{{\ttfamily
  1907.03780}}].

\bibitem{Luo:2019szz}
M.-x. Luo, T.-Z. Yang, H.~X. Zhu and Y.~J. Zhu, \emph{{Quark Transverse Parton
  Distribution at the Next-to-Next-to-Next-to-Leading Order}},
  \href{http://dx.doi.org/10.1103/PhysRevLett.124.092001}{\emph{Phys. Rev.
  Lett.} {\bfseries 124} (2020) 092001},
  [\href{https://arxiv.org/abs/1912.05778}{{\ttfamily 1912.05778}}].

\bibitem{Ebert:2020yqt}
M.~A. Ebert, B.~Mistlberger and G.~Vita, \emph{{Transverse momentum dependent
  PDFs at N$^3$LO}},  \href{https://arxiv.org/abs/2006.05329}{{\ttfamily
  2006.05329}}.

\bibitem{Scimemi:2019gge}
I.~Scimemi, A.~Tarasov and A.~Vladimirov, \emph{{Collinear matching for Sivers
  function at next-to-leading order}},
  \href{http://dx.doi.org/10.1007/JHEP05(2019)125}{\emph{JHEP} {\bfseries 05}
  (2019) 125}, [\href{https://arxiv.org/abs/1901.04519}{{\ttfamily
  1901.04519}}].

\bibitem{Jaffe:1996zw}
R.~L. Jaffe, \emph{{Spin, twist and hadron structure in deep inelastic
  processes}},  in \emph{{Ettore Majorana International School of Nucleon
  Structure: 1st Course: The Spin Structure of the Nucleon}}, pp.~42--129, 1,
  1996.
\newblock \href{https://arxiv.org/abs/hep-ph/9602236}{{\ttfamily
  hep-ph/9602236}}.

\bibitem{Braun:2003rp}
V.~Braun, G.~Korchemsky and D.~Müller, \emph{{The Uses of conformal symmetry
  in QCD}}, \href{http://dx.doi.org/10.1016/S0146-6410(03)90004-4}{\emph{Prog.
  Part. Nucl. Phys.} {\bfseries 51} (2003) 311--398},
  [\href{https://arxiv.org/abs/hep-ph/0306057}{{\ttfamily hep-ph/0306057}}].

\bibitem{Balitsky:1990ck}
I.~I. Balitsky and V.~M. Braun, \emph{{The Nonlocal operator expansion for
  inclusive particle production in e+ e- annihilation}},
  \href{http://dx.doi.org/10.1016/0550-3213(91)90618-8}{\emph{Nucl. Phys. B}
  {\bfseries 361} (1991) 93--140}.

\bibitem{Kanazawa:2013uia}
K.~Kanazawa and Y.~Koike, \emph{{Contribution of twist-3 fragmentation function
  to single transverse-spin asymmetry in semi-inclusive deep inelastic
  scattering}}, \href{http://dx.doi.org/10.1103/PhysRevD.88.074022}{\emph{Phys.
  Rev. D} {\bfseries 88} (2013) 074022},
  [\href{https://arxiv.org/abs/1309.1215}{{\ttfamily 1309.1215}}].

\bibitem{Echevarria:2015usa}
M.~G. Echevarria, I.~Scimemi and A.~Vladimirov, \emph{{Transverse momentum
  dependent fragmentation function at next-to--next-to--leading order}},
  \href{http://dx.doi.org/10.1103/PhysRevD.93.011502}{\emph{Phys. Rev. D}
  {\bfseries 93} (2016) 011502},
  [\href{https://arxiv.org/abs/1509.06392}{{\ttfamily 1509.06392}}].

\bibitem{Gamberg:2018fwy}
L.~Gamberg, Z.-B. Kang, D.~Pitonyak, M.~Schlegel and S.~Yoshida,
  \emph{{Polarized hyperon production in single-inclusive electron-positron
  annihilation at next-to-leading order}},
  \href{http://dx.doi.org/10.1007/JHEP01(2019)111}{\emph{JHEP} {\bfseries 01}
  (2019) 111}, [\href{https://arxiv.org/abs/1810.08645}{{\ttfamily
  1810.08645}}].

\bibitem{Mulders:1995dh}
P.~Mulders and R.~Tangerman, \emph{{The Complete tree level result up to order
  1/Q for polarized deep inelastic leptoproduction}},
  \href{http://dx.doi.org/10.1016/0550-3213(95)00632-X}{\emph{Nucl. Phys. B}
  {\bfseries 461} (1996) 197--237},
  [\href{https://arxiv.org/abs/hep-ph/9510301}{{\ttfamily hep-ph/9510301}}].

\bibitem{Braun:2009vc}
V.~Braun, A.~Manashov and J.~Rohrwild, \emph{{Renormalization of Twist-Four
  Operators in QCD}},
  \href{http://dx.doi.org/10.1016/j.nuclphysb.2009.10.005}{\emph{Nucl. Phys. B}
  {\bfseries 826} (2010) 235--293},
  [\href{https://arxiv.org/abs/0908.1684}{{\ttfamily 0908.1684}}].

\bibitem{Braun:2008ia}
V.~Braun, A.~Manashov and J.~Rohrwild, \emph{{Baryon Operators of Higher Twist
  in QCD and Nucleon Distribution Amplitudes}},
  \href{http://dx.doi.org/10.1016/j.nuclphysb.2008.08.012}{\emph{Nucl. Phys. B}
  {\bfseries 807} (2009) 89--137},
  [\href{https://arxiv.org/abs/0806.2531}{{\ttfamily 0806.2531}}].

\bibitem{Dreiner:2008tw}
H.~K. Dreiner, H.~E. Haber and S.~P. Martin, \emph{{Two-component spinor
  techniques and Feynman rules for quantum field theory and supersymmetry}},
  \href{http://dx.doi.org/10.1016/j.physrep.2010.05.002}{\emph{Phys. Rept.}
  {\bfseries 494} (2010) 1--196},
  [\href{https://arxiv.org/abs/0812.1594}{{\ttfamily 0812.1594}}].

\bibitem{Balitsky:1987bk}
I.~Balitsky and V.~M. Braun, \emph{{Evolution Equations for QCD String
  Operators}},
  \href{http://dx.doi.org/10.1016/0550-3213(89)90168-5}{\emph{Nucl. Phys. B}
  {\bfseries 311} (1989) 541--584}.

\bibitem{Bacchetta:2006tn}
A.~Bacchetta, M.~Diehl, K.~Goeke, A.~Metz, P.~J. Mulders and M.~Schlegel,
  \emph{{Semi-inclusive deep inelastic scattering at small transverse
  momentum}},
  \href{http://dx.doi.org/10.1088/1126-6708/2007/02/093}{\emph{JHEP} {\bfseries
  02} (2007) 093}, [\href{https://arxiv.org/abs/hep-ph/0611265}{{\ttfamily
  hep-ph/0611265}}].

\bibitem{Braun:2009mi}
V.~Braun, A.~Manashov and B.~Pirnay, \emph{{Scale dependence of twist-three
  contributions to single spin asymmetries}},
  \href{http://dx.doi.org/10.1103/PhysRevD.80.114002}{\emph{Phys. Rev. D}
  {\bfseries 80} (2009) 114002},
  [\href{https://arxiv.org/abs/0909.3410}{{\ttfamily 0909.3410}}].

\bibitem{Sohnius:1985qm}
M.~Sohnius, \emph{{Introducing Supersymmetry}},
  \href{http://dx.doi.org/10.1016/0370-1573(85)90023-7}{\emph{Phys. Rept.}
  {\bfseries 128} (1985) 39--204}.

\bibitem{Boer:2011xd}
D.~Boer, L.~Gamberg, B.~Musch and A.~Prokudin, \emph{{Bessel-Weighted
  Asymmetries in Semi Inclusive Deep Inelastic Scattering}},
  \href{http://dx.doi.org/10.1007/JHEP10(2011)021}{\emph{JHEP} {\bfseries 10}
  (2011) 021}, [\href{https://arxiv.org/abs/1107.5294}{{\ttfamily 1107.5294}}].

\bibitem{Scimemi:2018xaf}
I.~Scimemi and A.~Vladimirov, \emph{{Systematic analysis of double-scale
  evolution}}, \href{http://dx.doi.org/10.1007/JHEP08(2018)003}{\emph{JHEP}
  {\bfseries 08} (2018) 003},
  [\href{https://arxiv.org/abs/1803.11089}{{\ttfamily 1803.11089}}].

\bibitem{Geyer:1999uq}
B.~Geyer, M.~Lazar and D.~Robaschik, \emph{{Decomposition of nonlocal light
  cone operators into harmonic operators of definite twist}},
  \href{http://dx.doi.org/10.1016/S0550-3213(99)00334-X}{\emph{Nucl. Phys. B}
  {\bfseries 559} (1999) 339--377},
  [\href{https://arxiv.org/abs/hep-th/9901090}{{\ttfamily hep-th/9901090}}].

\bibitem{Belitsky:2000vx}
A.~V. Belitsky and D.~Mueller, \emph{{Twist- three effects in two photon
  processes}},
  \href{http://dx.doi.org/10.1016/S0550-3213(00)00542-3}{\emph{Nucl. Phys. B}
  {\bfseries 589} (2000) 611--630},
  [\href{https://arxiv.org/abs/hep-ph/0007031}{{\ttfamily hep-ph/0007031}}].

\bibitem{Ball:1998sk}
P.~Ball, V.~M. Braun, Y.~Koike and K.~Tanaka, \emph{{Higher twist distribution
  amplitudes of vector mesons in QCD: Formalism and twist - three
  distributions}},
  \href{http://dx.doi.org/10.1016/S0550-3213(98)00356-3}{\emph{Nucl. Phys. B}
  {\bfseries 529} (1998) 323--382},
  [\href{https://arxiv.org/abs/hep-ph/9802299}{{\ttfamily hep-ph/9802299}}].

\bibitem{Vladimirov:2017ksc}
A.~Vladimirov, \emph{{Structure of rapidity divergences in multi-parton
  scattering soft factors}},
  \href{http://dx.doi.org/10.1007/JHEP04(2018)045}{\emph{JHEP} {\bfseries 04}
  (2018) 045}, [\href{https://arxiv.org/abs/1707.07606}{{\ttfamily
  1707.07606}}].

\bibitem{Collins:2002kn}
J.~C. Collins, \emph{{Leading twist single transverse-spin asymmetries:
  Drell-Yan and deep inelastic scattering}},
  \href{http://dx.doi.org/10.1016/S0370-2693(02)01819-1}{\emph{Phys. Lett. B}
  {\bfseries 536} (2002) 43--48},
  [\href{https://arxiv.org/abs/hep-ph/0204004}{{\ttfamily hep-ph/0204004}}].

\bibitem{Vladimirov:2020umg}
A.~A. Vladimirov, \emph{{Self-contained definition of Collins-Soper kernel}},
  \href{https://arxiv.org/abs/2003.02288}{{\ttfamily 2003.02288}}.

\bibitem{Qiu:1991pp}
J.-w. Qiu and G.~F. Sterman, \emph{{Single transverse spin asymmetries}},
  \href{http://dx.doi.org/10.1103/PhysRevLett.67.2264}{\emph{Phys. Rev. Lett.}
  {\bfseries 67} (1991) 2264--2267}.

\bibitem{Scimemi:2016ffw}
I.~Scimemi and A.~Vladimirov, \emph{{Power corrections and renormalons in
  Transverse Momentum Distributions}},
  \href{http://dx.doi.org/10.1007/JHEP03(2017)002}{\emph{JHEP} {\bfseries 03}
  (2017) 002}, [\href{https://arxiv.org/abs/1609.06047}{{\ttfamily
  1609.06047}}].

\bibitem{Airapetian:2020zzo}
{\scshape HERMES} collaboration, A.~Airapetian et~al., \emph{{Azimuthal single-
  and double-spin asymmetries in semi-inclusive deep-inelastic lepton
  scattering by transversely polarized protons}},
  \href{https://arxiv.org/abs/2007.07755}{{\ttfamily 2007.07755}}.

\bibitem{Lefky:2014eia}
C.~Lefky and A.~Prokudin, \emph{{Extraction of the distribution function
  $h^{\perp}_{1T}$ from experimental data}},
  \href{http://dx.doi.org/10.1103/PhysRevD.91.034010}{\emph{Phys. Rev. D}
  {\bfseries 91} (2015) 034010},
  [\href{https://arxiv.org/abs/1411.0580}{{\ttfamily 1411.0580}}].

\bibitem{Chai:2018mwx}
X.~Chai, K.~Chen and J.~Ma, \emph{{A Note on Pretzelosity TMD Parton
  Distribution}},
  \href{http://dx.doi.org/10.1016/j.physletb.2018.12.020}{\emph{Phys. Lett. B}
  {\bfseries 789} (2019) 360--365},
  [\href{https://arxiv.org/abs/1808.10560}{{\ttfamily 1808.10560}}].

\bibitem{Ebert:2020gxr}
M.~A. Ebert, S.~T. Schindler, I.~W. Stewart and Y.~Zhao, \emph{{One-loop
  Matching for Spin-Dependent Quasi-TMDs}},
  \href{https://arxiv.org/abs/2004.14831}{{\ttfamily 2004.14831}}.

\bibitem{Vladimirov:2020ofp}
A.~A. Vladimirov and A.~Schäfer, \emph{{Transverse momentum dependent
  factorization for lattice observables}},
  \href{http://dx.doi.org/10.1103/PhysRevD.101.074517}{\emph{Phys. Rev. D}
  {\bfseries 101} (2020) 074517},
  [\href{https://arxiv.org/abs/2002.07527}{{\ttfamily 2002.07527}}].

\bibitem{Gribov:1972rt}
V.~Gribov and L.~Lipatov, \emph{{e+ e- pair annihilation and deep inelastic e p
  scattering in perturbation theory}}, {\emph{Sov. J. Nucl. Phys.} {\bfseries
  15} (1972) 675--684}.

\end{thebibliography}\endgroup

\end{document}